\documentclass[twocolumn]{aastex631}

\begin{document}

\title{Investigating stellar variability in the open cluster region NGC 381}

\author[0000-0001-5119-8983]{Maurya, Jayanand}
\affiliation{Aryabhatta Research Institute of observational sciencES (ARIES). Nainital, Uttrakhand, India}
\affiliation{School of Studies in Physics and Astrophysics, Pt. Ravishankar Shukla University, Raipur, Chattisgarh 492 010, India}
\affiliation{Astronomy \& Astrophysics Division, Physical Research Laboratory, Ahmedabad, 380009, State of Gujarat, India}
\author{Joshi, Yogesh C.}
\affiliation{Aryabhatta Research Institute of observational sciencES (ARIES). Nainital, Uttrakhand, India}

\author{Panchal, A.}
\affiliation{Aryabhatta Research Institute of observational sciencES (ARIES). Nainital, Uttrakhand, India}

\author{Gour, A. S.}
\affiliation{School of Studies in Physics and Astrophysics, Pt. Ravishankar Shukla University, Raipur, Chattisgarh 492 010, India}



\begin{abstract}

We study variable stars in the field of the open cluster NGC 381 using photometric data observed over 27 nights and identify a total of 57 variable stars out of which five are member stars. The variable stars are classified based on their periods, amplitudes, light curve shapes, and locations in the H-R diagram. We found a rich variety of variable stars in the cluster. We identified a total of 10 eclipsing binaries out of which 2 are Algol type (EA) while 8 are W UMa type (EW) binaries. The estimated ages of these EW binaries are greater than 0.6 Gyr which is in agreement with the formation time constraint of $\geqslant$ 0.6 Gyr on short-period eclipsing binaries. The estimation of the physical parameters of the three EW type binaries is done using PHOEBE model-fitting software. The pulsating variable stars include one each from $\delta$ Scuti and $\gamma$ Dor variability class. We determined the pulsation modes of pulsating variables with the help of the FAMIAS package. We obtained 15 rotational variables stars comprising four dwarf stars identified on the basis log(g) versus log(T$_{eff}$) diagram. These dwarf stars are found to have generally larger periods than the remaining rotational variables.

\end{abstract}

\keywords{stars: variable-open cluster:NGC 381-binaries:eclipsing -techniques: photometric}


\section{Introduction} \label{introduction}
Open clusters are known to host a rich variety of the variable stars \citep{2019MNRAS.482..658X,2020AJ....159...96S,2020MNRAS.499..618J}. These stars play a very crucial role in understanding the stars' intrinsic properties like pulsation, rotation, and emission. In addition to intrinsic properties, extrinsic properties of stars like geometrical effects in the case of eclipsing binaries can also be understood through variability study. Period-Luminosity relations of pulsating stars like Cepheids and RR Lyrae have been distinctly useful in the distance measurement and the Galactic structure study \citep{2020A&A...640A..92M,2019Sci...365..478S,2019MNRAS.482.3868I}. The intermediate-age clusters host a large number of $\delta$ Scuti  stars as their Main Sequence (MS) turn-offs lie close to the $\delta$ Scuti instability strip. $\delta$ Scuti-type variables are A$3$-F$0$ spectral class main-sequence or subgiant stars belonging to the lower classical instability strip. Intermediate age open clusters are the best observing targets for the asteroseismological study of $\delta$ Scuti-type stars \citep{2005PASP..117..601A}. $\gamma$ Dor stars are relatively a new class of main-sequence or sub-giant pulsating stars having spectral class A7-F5 and effective temperature T$_{eff}$ between 6700 to 7400 K. These stars are cooler than  $\delta$ Scuti stars and lie near the red edge of the $\delta$ Scuti instability strip. 

Eclipsing binary stars are quite common in open clusters as well. They are located above the single-star main sequence in the color-magnitude diagram of the clusters and can be used as a possible distance indicator \citep{2019Natur.567..200P} and in measurements of stellar masses and diameters \citep{2021AJ....161..221P}. Eclipsing binaries have also been useful in the calibration of theoretical stellar models \citep{2010A&ARv..18...67T}. The evolution from eclipsing binaries of Algol-type (EA) to W UMa type binaries (EW) provides an opportunity to understand angular momentum loss, tidally locked system, and mass transfer within binary systems \citep{2014MNRAS.438..859J,2020MNRAS.492.2731J,2020ApJS..249...18C}. The binary systems are also known to have pulsating companions like $\delta$ Scuti \citep{2017MNRAS.465.1181L,2021MNRAS.505.3206M}. The evolution of $\delta$ Scuti star in a binary system is affected by mass transfer and tidal interaction. The $\delta$ Scuti star in a binary system is very useful in theoretical modeling as the physical parameters of binary components can be accurately determined. Some eclipsing binaries exhibit asymmetry in the maxima of the light curves which is known as O' Connell effect due to stellar spots on the surface of the components \citep{2016RAA....16...63J}. 

The photometric variability is also caused by the modulation of the brightness due to rotation. Rotational variables belonging to the same cluster population are found to show a period-magnitude relation \citep{2020ApJS..246...15S}. These rotational variables have an uneven distribution of the cooler spots on the surface which produces variation in the magnitude due to the inclination angle of observation. The cooler spots and chromospheric activities generally happen in the late-type stars \citep{2021ApJS..253...51L}. \citet{2006AJ....131.1044D} found that rotation periods of chromospheric active stars decrease as the distance from the Galactic plane increases. This relation between period and distance from the Galactic plane was attributed to the fact that stars at low altitudes would be more evolved than stars at higher altitudes \citep{2006AJ....131.1044D}. Thus, the study of variable stars provides us a tool to investigate the interior of stellar evolution and the external properties of stars.

In the present study, we carried out a variability search and analysis of the open cluster NGC 381 for which the photometric study has earlier been carried out in \citet{2020MNRAS.494.4713M}. This is an intermediate-age sparse open cluster with an estimated age of 447 $\pm$ 52 Myr \citep{2020MNRAS.494.4713M}. It is located at a distance of 1131 $\pm$ 48 pc \citep{2020MNRAS.494.4713M}. The present variability study has been carried out in an $\sim$ 18 $\times$ 18 arcmin$^{2}$ region around the cluster center (RA = 01:08:19.57; DEC = +61:35:18.24). This is the very first variability study of the cluster NGC 381 which is part of our ongoing project of the variability investigations of the poorly studied open clusters \citep{2020MNRAS.499..618J}. In our previous study, we found a total of 132 member stars of the cluster after applying a 50$\%$ probability cut-off in this region. However, \citet{2018A&A...618A..93C} found a total of 148 member stars belonging to the cluster NGC 381 distributed over a much larger region than the observed region in the present study. The present paper is structured as follows. We discuss observations, data reduction, and the transformation of the instrumental magnitude to the standard magnitude in Section~\ref{observ}. The methods and techniques applied to detect variable stars in the present data set are described in Section~\ref{method}. We discussed the locations of the variable stars on the H-R diagram and their implications in Section~\ref{hr_diagram}. The classification and detailed analysis of the variable stars are discussed in Section~\ref{class}. We discuss and summarize our analysis in Section~\ref{discussion}.
\begin{table} 
\caption{The log for observed time series data of NGC 381 region. The 'L' and 'S' denote long and short exposures respectively.}
  \label{log}
  \centering
  \begin{tabular}{c c c c}  
  \hline  
  Date &  V filter  & I filter \\
       &  Number of frames & Number of frames \\
  \hline
  
 01/10/2017&  9L,  9S & 3L, 3S \\
 02/10/2017& 36L, 37S & 11L, 12S \\ 
 12/10/2017& 28L, 28S & 6L, 6S \\
 14/10/2017&  4L,  4S & 4L, 4S \\
 21/10/2017&  3L,  3S & 3L, 2S \\ 
 22/10/2017& 32L, 32S & - \\
 23/10/2017& 37L, 42S & 45L, 44S \\
 29/10/2017& 23L, 24S & 25L, 25S \\
 30/10/2017& 50L, 42S & - \\
 02/11/2017& 13L, 15S & 8L, 11S \\
 03/11/2017&  9L, 10S & 10L, 10S \\
 09/11/2017& 10L      & - \\
 12/11/2017&  6L,  6S & 17L, 17S \\
 14/11/2017&  6L,  6S & 7L, 7S \\
 16/11/2017& 24L, 21S & 26L, 26S \\
 17/11/2017&  4L,  4S & - \\
 24/11/2017&  8L      & - \\
 05/01/2018& 19L, 11S & 13L, 13S \\
 06/01/2018& 17L, 16S & 10L, 10S \\
 13/01/2018&  3L      & 3L \\
 22/01/2018&  3L      & 3L, 3S \\
 07/10/2018& 10L      & 3L \\
 20/10/2018&  7L      & - \\
 01/12/2018& 18L      & 19L \\
 04/12/2018& 21L, 12S & 12L, 12S \\
 07/12/2018& 8L       & - \\
 14/01/2019&  4L, 4S  & 5L, 5S \\
  \hline
  \end{tabular}
\end{table}

\section{Observations and Data analysis}\label{observ}
The observations for the current study were taken using 1.3-m Devasthal Fast Optical Telescope (DFOT). The DFOT telescope was mounted with a 2k$\times$2k CCD having wide coverage  of $\sim18^{'}$ $\times$ $18^{'}$ over the sky. The data for the variability analysis was collected during 27 nights spanning more than a year starting from 01 Oct. 2017. The observations were accumulated in 738 and 437 frames of the Johnson-Cousins V and I bands, respectively. We searched both long and short period variables using the collected data. The exposure times were varying between 180-220 and 80-120 sec for long exposures in V and I bands, respectively whereas short exposures for both the bands were in the range of 10-30 sec. The observation log for the observations of NGC 381 is given in Table~\ref{log}. 

The image cleaning procedures like bias correction, flat fielding, and cosmic rays removal were completed using suitable IRAF packages. We used DAOPHOT II packages and processes described by \citet{1987PASP...99..191S} to obtain instrumental magnitude through the point-spread function (PSF). The PSF technique is preferred for crowded fields like star clusters. The stars on the edges of some frames can contaminate the magnitudes of the adjacent stars and give a false sense of variability. We, therefore, removed all the data points within 20 pixels from the edges of each frame. We used DAOMATCH and DAOMASTER packages of DAOPHOT II to cross-match the same stars in different frames. The cluster NGC 381 was observed on 10 Oct. 2017 with Landolt's standard field SA 98 for the photometric calibration. The standardized data of NGC 381 was used in its photometric study \citep{2020MNRAS.494.4713M}. We converted the instrumental magnitude of the stars detected in frames of variability data to the standard magnitude by comparison with the standard data used in \citet{2020MNRAS.494.4713M}. The following conversion formula is used for obtaining the standard magnitudes of stars:
\begin{center}
V = a$_{1}$ $\times$ v + a$_{2}$ 
\end{center}
a$_{1}$ and a$_{2}$ used in the above relation are conversion coefficients. V is the standard magnitude while v is the instrumental magnitude of the same star. The conversion coefficients a$_{1}$ and a$_{2}$ are calculated through least-square fits for each frame of variability data. We did not use color terms in the conversion formula as the effect of the color terms was found to be very insignificant in the magnitude calibration. The I band data were reduced using the same methods as discussed for the reduction of V band data. However, the cluster was observed mostly in V-band. The number of I-band frames collected at any specific night were less than 20 for most of the nights. Due to more frames and coverage, V-band data was used for most of the analysis except for FAMIAS package. We used both V and I band data for the mode identification through the FAMIAS package which requires multi-band data. Only V-band observations were used for variability search. We produced light curves for a total of 5183 stars. All the light curves were inspected for the presence of variability.
\section{Identification}\label{method}
We used the Lomb-Scargle algorithm \citep{1976Ap&SS..39..447L,1982ApJ...263..835S} to search the periodic signal in the time series data. The Lomb-Scargle method is very efficient for periodicity search in unevenly spaced data. The period corresponding to the frequency of the maximum power in the power spectrum is generally used for phase calculation and phase-folding of the light curves. We checked all the periodic variables by phase folding the light curves for their periods. We chose only those variables which show good periodic brightness variation throughout the entire phase. 

We found a total of 57 periodic variables with periods ranging 41 minutes to 10.74 days in NGC 381 region. We verified the periods and the light curves of these periodic variables using \textit{NASA Exoplanet Archive Periodogram service} \citep{2013PASP..125..989A} and found that the two periods obtained were generally in agreement. Once the period of a star is known, we determined the phase of each epoch. We generated phase-folded light curves after binning in intervals of generally 0.01 phase. We calculated the mean magnitude for each phase bin. The binned light curves are useful in better identification of variables due to smoothing out of scattered points. We inspected the light curves of all the identified periodic variables for twice the estimated period to find primary and secondary minima.

We investigated the light curves of all the stars detected in the region. These variables were carefully examined for any blending by the neighboring stars and stars contaminated due to light from the neighboring star were discarded. The light curves having periods of 1/n day were excluded from the variability analysis to avoid false variability arising solely due to the aliasing effect. Here, n in the 1/n stands for a positive integer. The light curves were plotted after folding the data with phase. We examined the light curves of all the periodic variables for the presence of the primary and secondary minima as the signature of binary systems by plotting phase folded light curves corresponding to the twice of the obtained periods. The amplitude of the variation in periodic variables was calculated by taking half of the difference in the mean magnitudes of the three faintest and the three brightest points in the curves \citep{2020MNRAS.499..618J}. We have provided the basic parameters like period, amplitude, and magnitude in Table~\ref{basic_par}. 
\begin{table*}\fontsize{6.8}{6.8}\selectfont
\caption{The basic parameters of the identified variable stars. The columns denote star id, RA, DEC, V magnitude, period, amplitude, T$_{eff}$, log (T$_{eff}$), bolometric magnitude, luminosity, membership, and types of variability. 'M' indicates cluster member while 'F' marks field star.}
\label{basic_par}
\centering
\begin{tabular}{cccccccccccc}
\hline
 ID   &  RA (J2000) & DEC (J2000)  &   V     &   Period   & $\Delta$V &  T$_{eff}$&   log(T$_{eff}$) & M$_{bol}$& log(L/L$_{\odot}$) & Membership & Type \\
      & (hh:mm:ss)  & (dd:mm:ss)   & (mag)   &   (day)    &    (mag)   &    (K)   &  (K)  &    &    &     \\
\hline
     199&    17.38816&    61.47656& 19.659& 0.64170& 0.117& 6058& 3.78234& 5.138103&-0.163241& F& Misc\\
     298&    17.37635&    61.51619& 14.970& 2.88585& 0.274& 7156& 3.74758& 1.067196& 1.465122& F& EA \\
     528&    17.34962&    61.53601& 19.578& 0.63944& 0.240& 5866& 3.76832& 6.143452&-0.565381& F& Misc\\
     535&    17.34960&    61.46318& 18.186& 0.88731& 0.103& 6708& 3.82659& 3.906007& 0.329597& F& Misc\\
     600&    17.34102&    61.63476& 17.804& 0.30660& 0.071& 4873& 3.68779& 0.307142& 1.769143& F& EW \\
     608&    17.34195&    61.47071& 19.597& 0.20628& 0.104& 5649& 3.75199& 5.342264&-0.244906& F& Rot \\
     610&    17.34198&    61.44597& 18.093& 0.88574& 0.095& 6352& 3.80293& 3.072912& 0.662835& F& Misc\\
     732&    17.32448&    61.71775& 19.529& 0.30934& 0.259& 6642& 3.82227& 4.805819&-0.030327& F& EW \\
     737&    17.32681&    61.43044& 16.723& 5.90783& 0.068& 7300& 3.86335& 1.245203& 1.393919& F&  Misc\\
     749&    17.32584&    61.44954& 16.834& 0.85538& 0.061& 4331& 3.62835& 7.357904&-1.051162& F& Rot \\
    1155&    17.27404&    61.67363& 18.785& 0.30390& 0.056& 5382& 3.73098& 5.404430&-0.269772& F& Rot \\
    1173&    17.27376&    61.50248& 16.161& 7.97406& 0.042& 5058& 3.70359& 0.672518& 1.622993& F& Rot \\
    1217&    17.26697&    61.65434& 16.034& 1.70109& 0.037& 6271& 3.71063& 3.371889& 0.543244& F& Rot \\
    1397&    17.25031&    61.47081&  9.858& 2.48932& 0.265& 8059& 4.19605& -4.264955& 3.597982& F& EA \\
    1399&    17.25022&    61.43972& 14.344&15.60373& 0.065& 6029& 3.76993& 4.701243& 0.011503& F& Rot\\
    1441&    17.24511&    61.51020& 16.333& 2.28997& 0.064& 5007& 3.69958& 0.278268& 1.780693& F& Misc\\
    1596&    17.22836&    61.60860& 14.118& 0.43500& 0.022& 7158& 3.82861& 3.137557& 0.636977& M& $\gamma$ Dor \\
    1602&    17.22763&    61.58761& 15.413& 1.13352& 0.030& 6922& 3.80548& 3.324183& 0.562327& F& Misc \\
    1689&    17.21949&    61.44122& 19.499& 0.85503& 0.120& 5739& 3.75883& 4.550551& 0.071779& F& Misc\\
    1785&    17.20845&    61.56744& 17.962& 0.47661& 0.207& 5841& 3.76648& 3.387451& 0.537020& F& EW \\
    1812&    17.20693&    61.43172& 17.971& 0.46948& 0.080& 5075& 3.70547& 3.594283& 0.454287& F& Rot \\
    1935&    17.19160&    61.43677& 14.743&18.86972& 0.059& 4966& 3.71715& 5.490398&-0.304159& F& Misc\\
    2018&    17.17960&    61.61100& 20.054& 0.38967& 0.153& 4119& 3.61478& 7.402637&-1.069055& F& Misc\\
    2058&    17.17713&    61.43163& 18.790& 0.46970& 0.107& 6259& 3.79651& 3.889816& 0.336073& F& Misc\\
    2092&    17.17362&    61.44512& 16.624& 0.46971& 0.065& 6489& 3.80542& 3.705286& 0.409886& F& Misc\\
    2245&    17.15525&    61.64929& 15.623& 0.96225& 0.023& 7869& 3.83518& 2.063115& 1.066754& F& Misc\\
    2302&    17.14846&    61.71435& 13.743& 1.46727& 0.042& 5976& 3.77641& 4.751433&-0.008573& F& Rot\\
    2458&    17.13336&    61.67006& 19.243& 0.27405& 0.106& 5012& 3.69791& 6.883659&-0.861464& F& EW \\
    2472&    17.13248&    61.60869& 13.045& 1.27248& 0.026& 7574& 4.02109& 0.742912& 1.594835& M& Misc\\
    2483&    17.13197&    61.57956& 15.163& 0.07504& 0.037& 6731& 3.81758& 1.983478& 1.098609& F& $\delta$-Scuti\\
    2597&    17.11615&    61.71983& 19.354& 0.76046& 0.110& 5804& 3.76370& 4.313651& 0.166540& F& Misc\\
    2618&    17.11641&    61.58124& 17.374& 0.24802& 0.020& 4825& 3.68353& 0.905611& 1.529756& F& Rot \\
    2673&    17.10656&    61.67836& 11.645&12.19928& 0.042& 7956& 3.88981& 0.749634& 1.592147& M& Misc\\
    2699&    17.10342&    61.68298& 19.325& 0.30427& 0.081& 5658& 3.75265& 5.028546&-0.119419& F& Rot \\
    2999&    17.07245&    61.51315& 15.389& 2.93362& 0.035& 6516& 3.79020& 4.144697& 0.234121& M& Rot\\
    3005&    17.07080&    61.60435& 19.634& 0.31405& 0.150& 4802& 3.68145& 5.670018&-0.376007& F& EW\\
    3472&    17.02022&    61.49097& 19.207& 0.38882& 0.126& 5653& 3.75225& 5.138588&-0.163435& F& EW \\
    3535&    17.01155&    61.57847& 17.657& 0.32339& 0.112& 5784& 3.70447& 5.917628&-0.475051& F& EW \\
    3680&    16.99572&    61.62203& 19.166& 0.84519& 0.216& 7238& 3.85964& 3.430055& 0.519978& F& Misc\\
    3717&    16.99284&    61.54328& 18.603& 0.58746& 0.089& 4873& 3.68783& 5.478819&-0.299528& F& Rot\\
    3792&    16.98570&    61.44255& 15.835& 0.68694& 0.049& 4795& 3.65312& 6.642276&-0.764910& F& Rot\\
    4049&    16.95194&    61.57502& 13.343& 2.94373& 0.024& 7988& 3.90244&-0.484138& 2.085655& F& Misc\\ 
    4070&    16.95224&    61.44262& 17.795& 0.31974& 0.046& 5273& 3.72202& 1.591900& 1.255240& F& Misc\\
    4243&    16.92789&    61.53868& 17.463& 2.13002& 0.084& 5615& 3.77448& 4.823012&-0.037205& F& Misc\\
    4281&    16.91943&    61.67900& 18.635& 0.24194& 0.054& 7211& 3.85798& 3.439219& 0.516313& F& Misc\\
    4283&    16.92179&    61.55971& 19.408& 0.84517& 0.131& 6280& 3.79796& 4.464289& 0.106285& F& Misc\\
    4361&    16.90875&    61.60868& 15.017& 0.40748& 0.024& 6392& 3.81621& 3.751129& 0.391548& M& Misc\\
    4405&    16.90401&    61.55529& 16.117& 0.76060& 0.032& 5239& 3.83875& 2.403936& 0.834332& F& Rot\\
    4525&    16.88284&    61.72299& 20.046& 0.26657& 0.158& 4546& 3.65762& 5.943748&-0.485499& F& Misc\\
    4705&    16.86359&    61.57545& 18.034& 0.55897& 0.076& 5209& 3.70465& 3.444240& 0.514304& F& Misc\\
    4829&    16.84602&    61.71936& 19.438& 0.34551& 0.114& 4409& 3.64436& 3.735335& 0.397866& F& EW\\
    5091&    16.81309&    61.56333& 14.800& 0.78344& 0.062& 6335& 3.80175& 3.469250& 0.504300& M& Misc\\
    5166&    16.80137&    61.66182& 16.948& 0.89365& 0.065& 6170& 3.80247& 3.381778& 0.539289& F& Misc\\
    5199&    16.79518&    61.70208& 18.722& 1.13805& 0.126& 6601& 3.81958& 3.488482& 0.496607& F& Misc\\
    5370&    16.77350&    61.65828& 19.460& 1.14294& 0.142& 5029& 3.70151& 6.353474&-0.649389& F& Misc\\ 
    5423&    16.77012&    61.50882& 15.765& 0.46019& 0.032& 4423& 3.64572&-0.725947& 2.182379& F& Rot\\
    5451&    16.76296&    61.61349& 16.935& 1.13334& 0.054&10619& 4.02610& 0.287079& 1.777169& F& Misc\\ 
\hline
\end{tabular}
\end{table*}

\begin{figure}
\centering
\includegraphics[width=9 cm, height=7 cm]{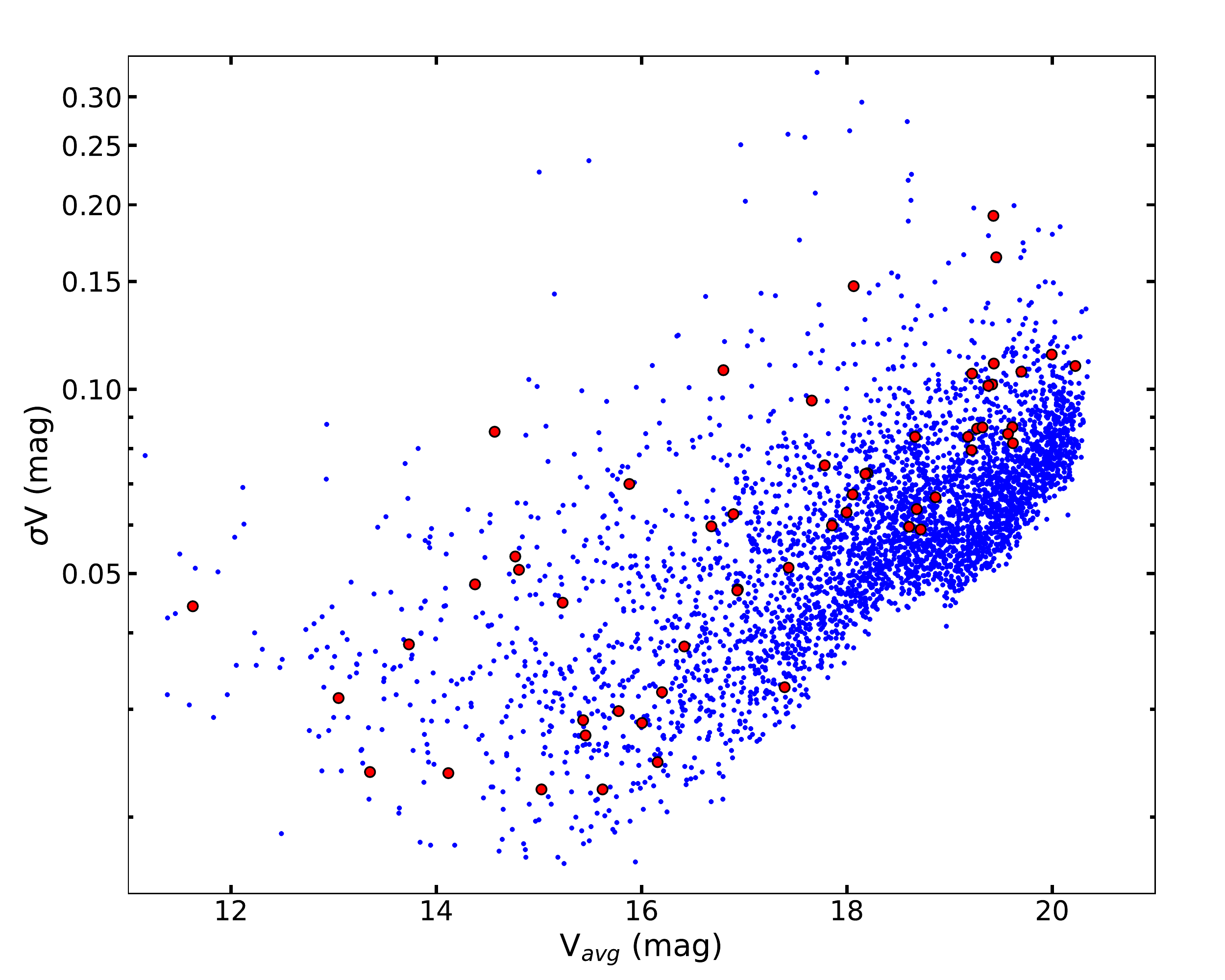}
\caption{The light curve RMS as a function of the magnitudes for the stars in the NGC 381 region. The y-axis in the figure is on the logarithmic scale. The red larger points denotes the variable stars whereas blue points in the background are for the non-variable stars.}
\label{rms}
\end{figure}

\begin{figure*}
\includegraphics[width=18 cm, height=14 cm]{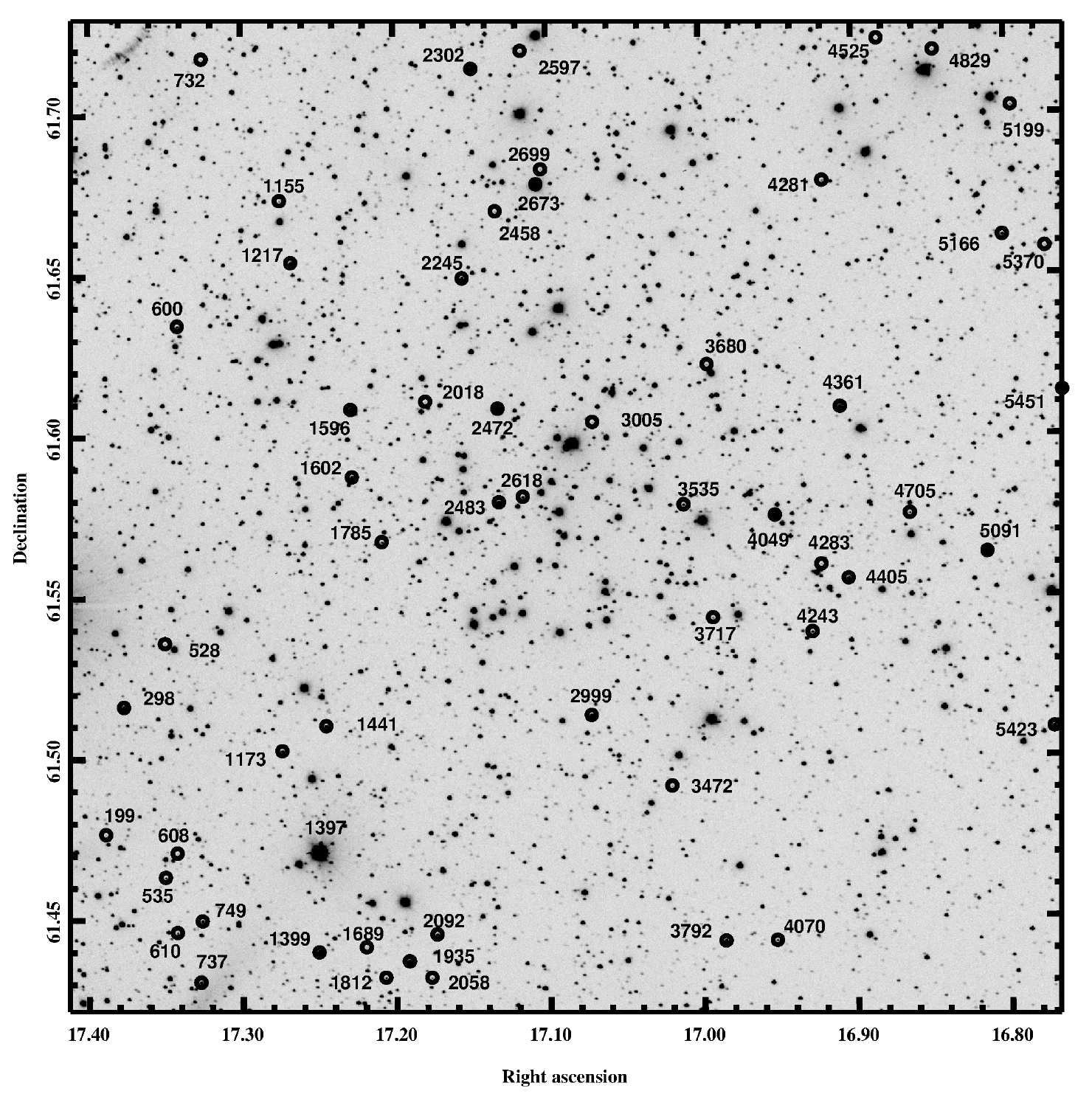}
\caption{The finding chart for 57 periodic variables detected in NGC 381 region. The variable stars are marked by the circle and labeled with star IDs. The right ascension and declination are given in degrees.}
\label{fchart}
\end{figure*}

\begin{figure*}
\includegraphics[width=18 cm, height=17 cm]{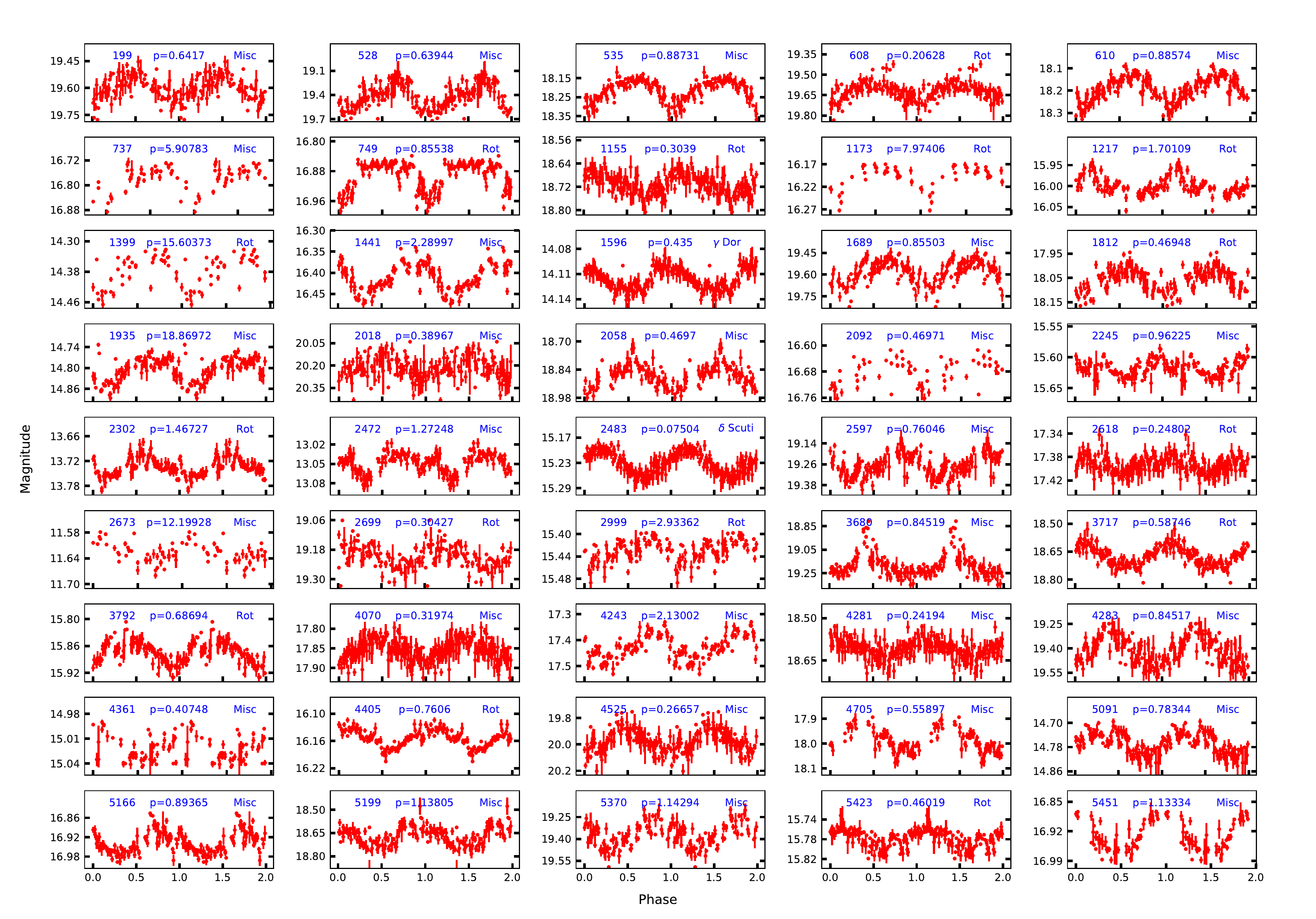}
\caption{The phase folded light curves of periodic variable stars identified in the NGC 381 region. The plots are given for twice of the phase value for a good presentation. The star IDs, periods (p), and variability types are denoted at the top of each light curve. The original light curves and Lomb-Scargle periodograms are shown in the appendix of this paper. The light curves for the eclipsing binaries are given in Figure~\ref{lc_eb}. The light curves for stars 1602 and 4049 are shown in Figure~\ref{rscvn}. These two stars are reported to be RS Canum Venaticorum-type (RS CVn) variables in the literature \citep{2006SASS...25...47W}.}
\label{lc_var}
\end{figure*}

We found a total of 57 periodic variables after all the mentioned examinations. We have determined the light curve root mean square (RMS) to investigate the spread in data points for variable stars.  The plot for the light curves RMS against the mean magnitudes of the stars is shown in Figure~\ref{rms}. It is evident in  the figure that the variable stars are generally associated with larger standard deviation in the magnitudes compared to non-variable stars. The finding chart for all the 57 variable stars is given in Figure~\ref{fchart}. 
The obtained periodic variables include 39 short period (p$<$1 day) and 18 long period (p$>$1 day) variables. The mean magnitudes of these variable stars vary from 9.858 mag to 20.054 mag in the V band. The estimated periods of these variables are in the range of $\sim$3 hours to $\sim$19 days while amplitude variation is between 0.022 to 0.256 mag. We found 10 out of 57 periodic variables as eclipsing binary stars. The light curves of the variables other than eclipsing binaries and RS CVn stars are shown in Figure~\ref{lc_var}. Six stars out of these 57 variables are member stars as mentioned in Table~\ref{basic_par}. The 51 remaining variable stars belong to the field population found in the cluster NGC 381 region.
\section{Variable stars and H-R diagrams}\label{hr_diagram}
The H-R diagram basically represents the empirical relation between spectral type and luminosity. The other forms of the H-R diagram can be temperature versus luminosity or color index versus absolute magnitude. The location of a star on the H-R diagram is affected by the factors like stellar wind, magnetic field, rotation, and chemical abundance besides initial mass and age. The positions of the variable stars on the H-R diagrams provide us with information about the evolutionary stage of the stars. The H-R diagrams have also been immensely useful in classifying the variable stars as they offer constrained regions for different classes of variable stars.
\subsection{Color-magnitude diagram}
\begin{figure*}
\centering
\includegraphics[width=16 cm, height=11 cm]{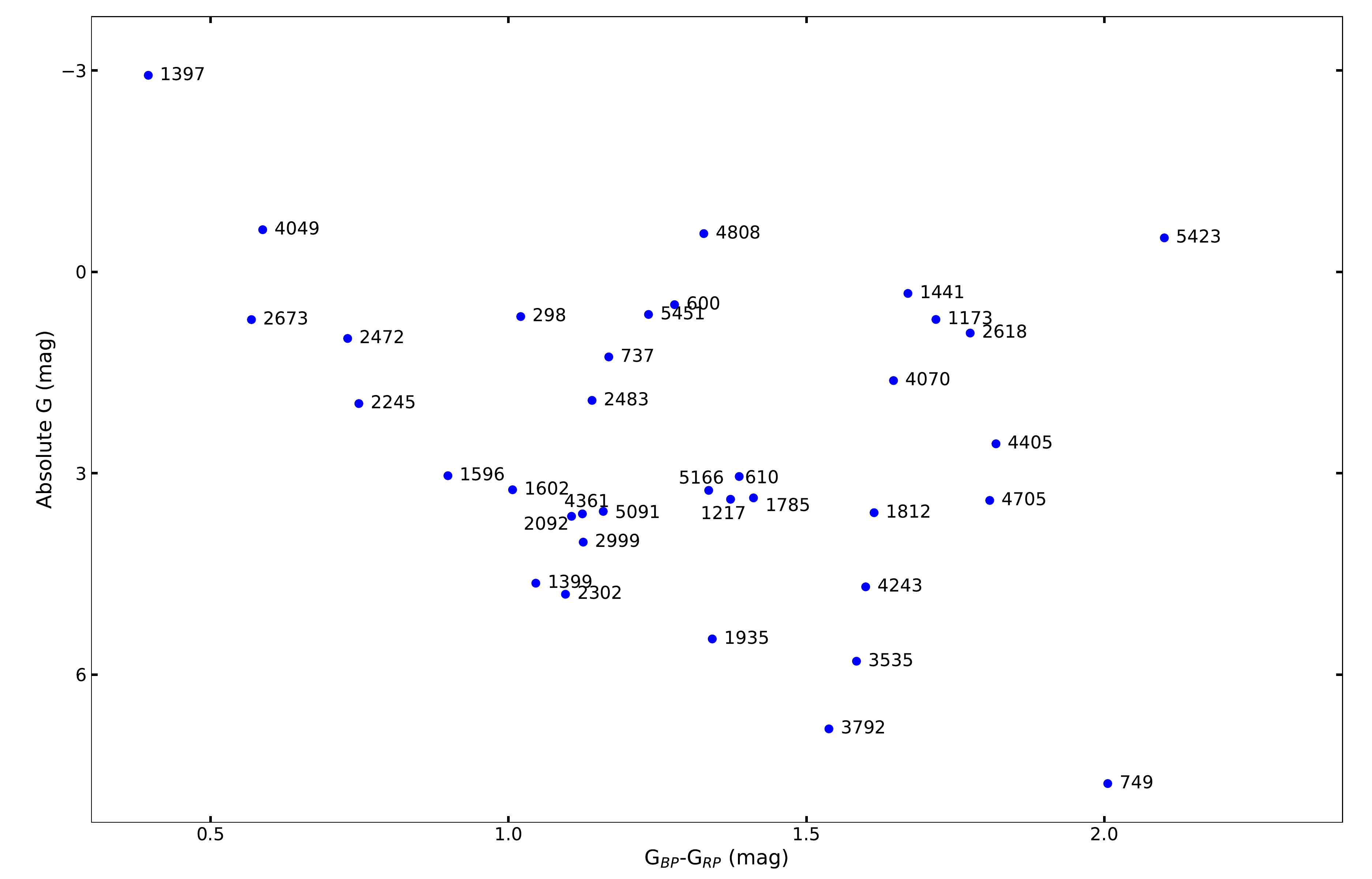}
\caption{Plot of absolute-G versus (G$_{BP}$ - G$_{RP}$) color-magnitude diagram for the 35 variable stars identified in the NGC 381 region.}
\label{CMD}
\end{figure*}

The absolute G versus (G$_{BP}$ - G$_{RP}$) color-magnitude diagram (CMD) has been useful in the classification of variable stars. The CMD is particularly important as it provides constrained regions for variable types like eclipsing binaries, rotational, eruptive, and cataclysmic variables besides pulsating variables \citep{2019A&A...623A.110G,2021AJ....162..133S}. The pulsating variables like $\delta$-Scuti and $\gamma$ Dor have very localized clumps on CMD. The separate locations of variables on the CMD help in removing ambiguity due to the similarity in light curves of two different classes of variables \citep{2021AJ....162..133S}. We used absolute G magnitudes, G$_{BP}$, and G$_{RP}$ band data for 35 variables provided by \citet{2019A&A...628A..94A}. In the study, the absolute G magnitudes were calculated using precise Gaia DR2 parallaxes. These magnitudes for the variable stars are given in Table~\ref{abs_mag}. We have shown the CMD constructed from these 35 variable stars in Figure~\ref{CMD}. 
\begin{table}\fontsize{7.2}{7.2}\selectfont
\caption{G, G$_{BP}$, and G$_{RP}$ magnitudes are taken from \textit{Gaia} eDR3 mag. The absolute G magnitudes denoted by G$_{0}$ are used from \citet{2019A&A...628A..94A}. The first column gives star IDs.}
 \label{abs_mag}
 \centering
 \begin{tabular}{cccccc}
 \hline
 ID   &  G  &  G$_{BP}$&  G$_{RP}$&  (G$_{BP}$-G$_{RP}$)&  G$_{0}$   \\
      &(mag)&     (mag)& (mag)    &  (mag) &    \\
 \hline
     298& 14.3877&   14.7998&   13.7793&   1.020500& 0.666100 \\
     600& 17.5191&   18.0799&   16.8012&   1.278700& 0.488836 \\
     610& 17.7972&   18.4208&   17.0336&   1.387200& 3.048655 \\
     737& 16.5466&   17.0495&   15.8812&   1.168300& 1.267372 \\
     749& 16.1014&   17.1122&   15.1064&   2.005800& 7.622625 \\
    1173& 15.6201&   16.4421&   14.7247&   1.717400& 0.709279 \\
    1217& 15.7076&   16.2976&   14.9249&   1.372700& 3.387695 \\
    1397& 9.8681&    10.002&     9.6066&   0.395400&-2.927836 \\
    1399& 14.1543&   14.5896&   13.5438&   1.045800& 4.637926 \\
    1441& 15.8335&   16.6254&    14.955&   1.670400& 0.321563 \\
    1596& 13.9532&   14.3141&   13.4158&   0.898300& 3.036791 \\
    1602& 15.2371&   15.6453&   14.6385&   1.006800& 3.246058 \\
    1785& 17.6741&   18.3018&   16.8906&   1.411200& 3.367079 \\
    1812& 17.5073&   18.2624&   16.6488&   1.613600& 3.587901 \\
    1935& 14.4212&   15.0251&   13.6831&   1.342000& 5.468198 \\
    2092& 16.4239&   16.8871&   15.7813&   1.105800& 3.641529 \\
    2245& 15.5027&   15.7892&   15.0404&   0.748800& 1.96228 \\
    2302& 13.9527&   13.9487&   12.8531&   1.095600& 4.802494 \\
    2472& 12.9203&   13.2031&   12.4731&   0.730000& 0.992314 \\
    2483& 14.9365&    15.416&   14.2758&   1.140200& 1.914954 \\
    2618& 16.7424&   17.5992&   15.8243&   1.774900& 0.91183 \\
    2673& 11.5709&   11.7849&   11.2164&   0.568500& 0.711613 \\
    2999& 15.1195&   15.5919&   14.4665&   1.125400& 4.026561 \\
    3535& 17.0551&   17.7956&   16.2115&   1.584100& 5.800122 \\
    3792& 15.3578&   16.0753&   14.5375&   1.537800& 6.807869 \\
    4049& 13.2593&   13.4677&   12.8803&   0.587400&-0.627663 \\
    4070& 17.3223&    18.103&   16.4568&   1.646200& 1.619608 \\
    4243& 16.9251&   17.6806&   16.0811&   1.599500& 4.692011 \\
    4361& 14.7457&   15.2224&   14.0984&   1.124000& 3.604831 \\
    4405& 15.5044&   16.3757&   14.5575&   1.818200& 2.561388 \\
    4705& 17.3425&    18.225&   16.4173&   1.807700& 3.404837 \\
    5091& 14.7421&   15.1451&   13.9861&   1.159000& 3.56836 \\
    5166& 16.5781&   17.1719&   15.8359&   1.336000& 3.255291 \\
    5423& 14.9632&   16.0395&   13.9387&   2.100800&-0.506022 \\
    5451& 16.661 &   17.1892&   15.9542&   1.235000& 0.634745 \\
 \hline
 \end{tabular}
\end{table}
%

The CMD for variable stars in the present study indicates the absence of stars like the long-period pulsating variable, $\alpha$ Cygni, R Coronae Borealis, V361 Hydrae, and ZZ Ceti stars as constrained regions for these types of stars lie outside the current CMD \citep{2019A&A...623A.110G}. We have considered the position of variable stars on the CMD during their classification. 
\subsection{H-R diagram}
%
\begin{figure*}
\includegraphics[width=17 cm, height=16 cm]{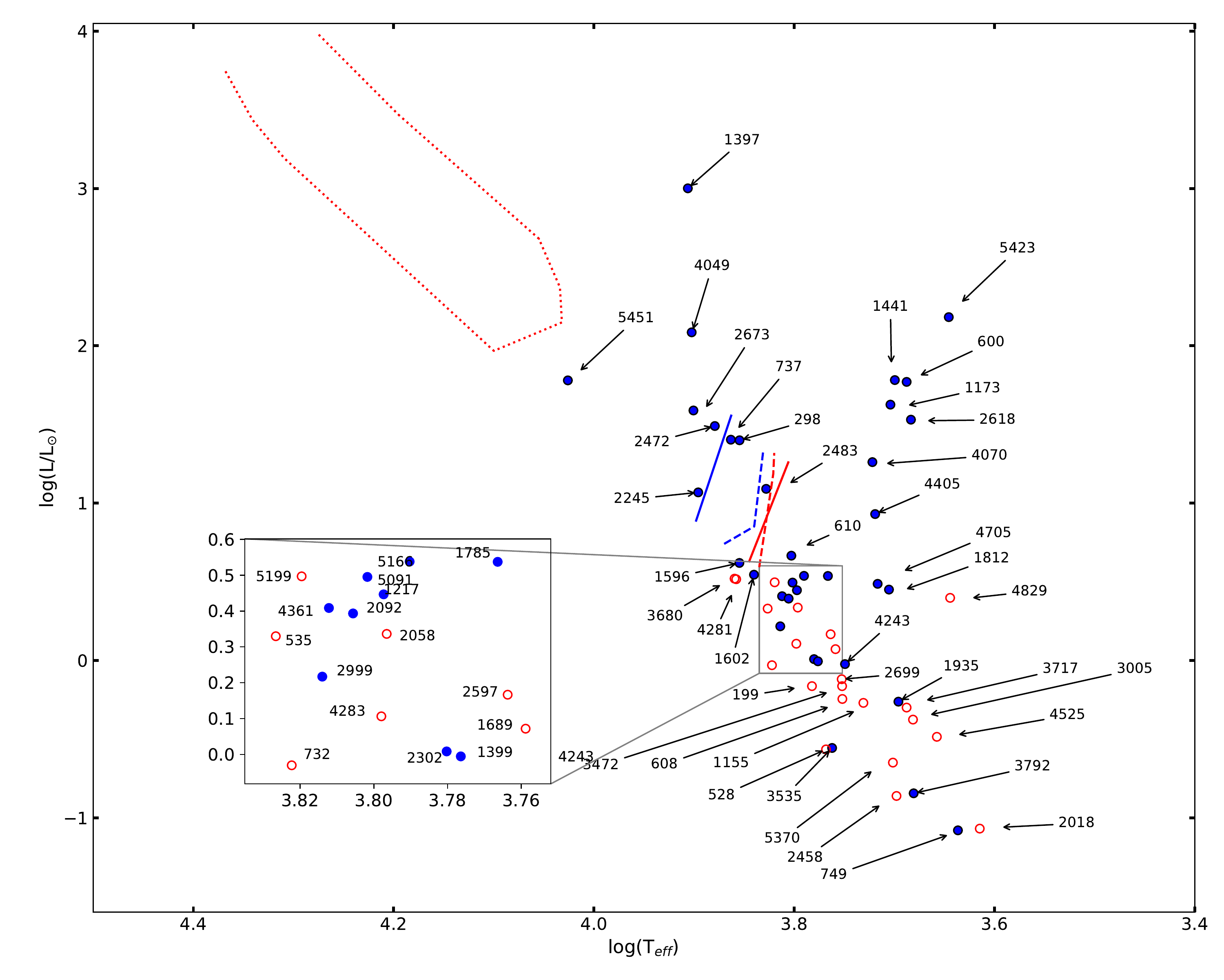}
\caption{H-R diagram of variable stars found in NGC 381 region. The instability strips corresponding to slowly pulsating B type, $\delta$ Scuti, and $\gamma$ Dor stars are shown by red dotted curve, continuous line, and dashed line, respectively. The blue filled points denote the stars which were available in \citet{2019A&A...628A..94A} and \citet{2019AJ....158...93B} catalogs. The open circles represent stars whose effective temperatures were calculated using \citet{2010AJ....140.1158T} relations.}
\label{hr_n381}
\end{figure*}
We examined log(L/L$_{\odot}$) versus log(T$_{\rm eff}$) H-R diagram in the present study to know the evolutionary stage of the identified variable stars. The effective temperature, T$_{eff}$,  of 24 stars were available in the catalog provided by \citet{2019AJ....158...93B}. These temperatures were calculated using a large training sample for regression of about four million stars which makes their estimated effective temperatures more precise than the effective temperatures of Gaia DR2. We used effective temperatures from \citet{2019A&A...628A..94A} catalog for additional 11 stars. The log(T$_{eff}$) for the remaining 22 variable stars were calculated from (B-V)$_{0}$ color through \citet{2010AJ....140.1158T} relations using distances from \citet{2018AJ....156...58B} and reddening from 3D reddening map given by \citet{2019ApJ...887...93G} which provides reddening E(g-r) with very good resolution for given right ascension, declination, and distance. We calculated E(B-V) from E(g-r) using extinction ratio relations given by \citet{2019ApJ...877..116W}. We, then, calculated (B-V)$_{0}$ values from (B-V) color of the star using E(B-V). These values of log(T$_{eff}$) were used in calculation of logarithmic luminosity, log(L/L$_{\odot}$), using the method described by \citet{2010AJ....140.1158T}. The logarithmic luminosity values were calculated using the following equations: 
\begin{center}
M$_{bol}$ = M$_{v}$+BC$_{v}$ \\
log(L/L$_{\odot}$) = -0.4(M$_{bol}$-M$_{bol_{\odot}}$) \\
\end{center}
The bolometric magnitude M$_{v}$ is calculated from magnitude in V band, distance, and A$_{v}$ extinction. The values of distance and extinction were taken from \citet{2019A&A...628A..94A} which were estimated with \textit{Gaia} DR2 data. The bolometric correction (BC$_{v}$) values were calculated using empirical relations given by \citet{2010AJ....140.1158T}. The M$_{bol_{\odot}}$ in the above equations is the bolometric magnitude of the Sun which is taken as 4.73 mag \citep{2010AJ....140.1158T}. Once we had calculated log(T$_{eff}$), the values of log(L/L$_{\odot}$) were calculated using the above described relations through \citet{2010AJ....140.1158T} method of BC$_{v}$ calculation. We plotted the H-R diagram using the calculated log(T$_{eff}$) and log(L/L$_{\odot}$) as shown in Figure~\ref{hr_n381}. The instability strips corresponding to $\delta$-Scuti, $\gamma$ Dor, SPB, and Cephei stars are also marked in the H-R diagram. 

The instability strips belonging to high mass stars like Slowly Pulsating B type (SPB) stars do not host any star in the H-R diagram of NGC 381. As most of the variable stars in the NGC 381 region belong to the field population, the lack of high mass variable stars is along with the expectations. We found three variable stars in the gap between SPB and $\delta$-Scuti strips which is theoretically inhibited for pulsation. The variable stars in this gap have also been found in previous studies \citep {2011MNRAS.413.2403B,2020MNRAS.492.3602J} and we have discussed this phenomenon in Section~\ref{class}. There are few variable stars in the instability strips for low mass variables like $\delta$-Scuti and $\gamma$ Dor. The locations of variable stars in the H-R diagram as well as CMD have been used for the classification of variables. In the following section, we individually describe the characteristics of the different classes of variables found in our study.
%
\section{Classification}\label{class}
The classification of variable stars is necessary for their comparative study. The variable stars have been classified based on period, amplitude, effective temperature, mass, the shape of the light curves, and location on the H-R diagram. Pulsating variable stars are known to be located in various instability strips related to the pulsating mechanism on the H-R diagram. Additionally, the locations of the variable stars on the CMD diagram were also considered for the classification of the variability class. Despite of our best efforts, we acknowledge that there can be some ambiguity in the classifications of the variable stars for instance the rotational variable can also show primary and secondary minima. Similarly, there can be ambiguity in the classification of the  $\delta$-Scuti and $\gamma$ Dor stars as they belong to closely placed instability strips on the H-R diagram. The gaps in the data can also affect the shape of the light curves especially for the low amplitude variables.
\subsection{Variability caused by Eclipses and Rotation}
\subsubsection{Eclipsing Binaries}
Eclipsing binaries have been greatly useful in the precise determination of masses, luminosities, radii, and distances \citep{2019Natur.567..200P,2021ApJ...907L..33S}. We identified a number of eclipsing binaries from the inspection of the light curves. The types of photometric eclipsing binaries are identified through specific features like the size of the primary and secondary minima in the light curves. These specific features can be theoretically explained by Roche lobes geometry. In eclipsing binary of Algol type (EA), one of the component stars has filled or overflowed the Roche lobe. The light curves of these variables are identified by the clear beginning and end of the maxima and minima. The light curves of eclipsing binaries of $\beta$-Lyrae type (EB) do not show any definite start or end of eclipses as light continuously varies. The EBs are semi-detached systems having gravitationally distorted components and hence have continuously varying light curves. The W Ursae Majoris-type (EW) eclipsing binaries or W UMa stars are marked by almost equal primary and secondary minima in the light curves. EW binaries are contact binaries consisting of ellipsoidal components with periods generally less than 1 day and no clear onset or end of the eclipses. It has been found that EA binaries evolve to EW through angular momentum loss and mass transfer \citep{2018ApJS..235....5Q}.

\begin{figure*}
\centering
\includegraphics[width=18 cm, height=12 cm]{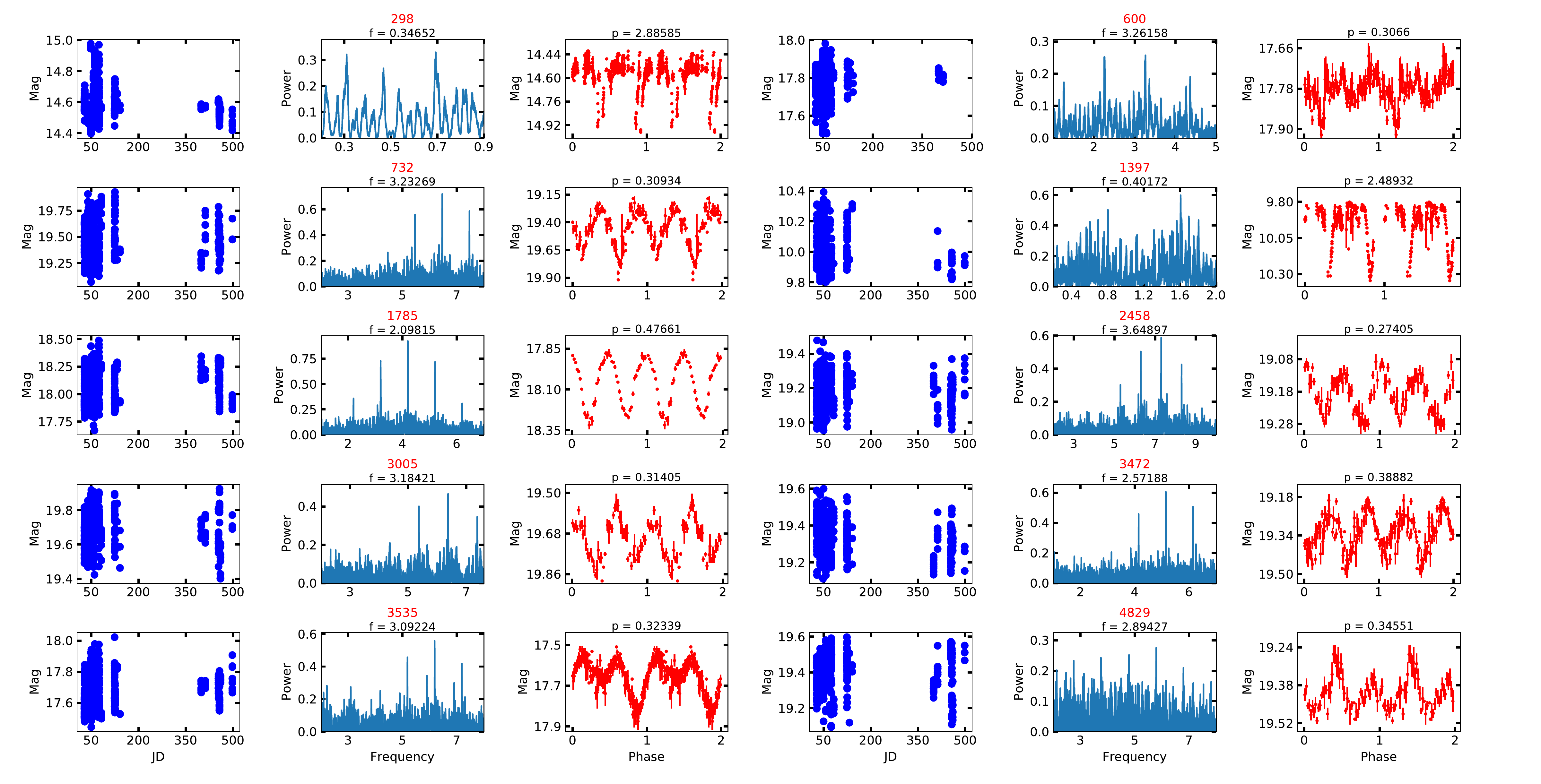}
\caption{The original light curves (left subplot), Lomb-Scargle periodograms (middle subplot), and phase folded  light curves  (right subplots) of each eclipsing binary star found in the NGC 381 region. The periods (p) and corresponding frequencies (f) are also shown in the subplots. Star IDs are given at the top of each plot.}
\label{lc_eb}
\end{figure*}

We found two EA and eight EW binaries in NGC 381 region. The high number of EW binaries in the NGC 381 region compared to EA binaries is consistent with the fact that a study by \citet{2020ApJS..249...18C} detected EW binaries many times more than EA binaries. The light curves for obtained eclipsing binaries are given in Figure~\ref{lc_eb}. The light curves of binary stars 298 and 1397 are marked with clear beginning and end of primary secondary minima. Though all three types of eclipsing binaries have wide-spread locations on CMD, however, EA-type binaries have been found to be located at the most extreme than the other types. The binary star ID 1397 is located on the CMD in the region mostly occupied by EA-type binaries \citep[e.g.,][]{2019A&A...623A.110G}. Therefore we classified star 1397 as an EA-type eclipsing binary. The light curve properties of binary star 298 suggest it to be an EA-type eclipsing binary which was also supported by its location on the CMD. The light curves of the remaining 8 eclipsing binaries having IDs 600, 732, 1785, 2458, 3005, 3472, 3535, and 4829 show characteristics of EW-type binaries. The stars having ID 600, 1785, and 3535 are also shown on the CMD and their locations on the CMD are consistent with their classification as EW stars. The positions of the EA binary stars 298 and 1397 on the CMD are significantly higher than the locations of EW binary systems indicating primary stars of earlier spectral type in the case of EA binaries compared to primary stars of EW systems. We found periods of EW-type binaries to be significantly smaller than the EA-type binaries as expected by theories of EA to EW evolution \citep{2020ApJS..249...18C}. The periods of all these binaries are found to be greater than the observed period cut-off of 0.23 days for eclipsing binaries \citep{2007MNRAS.382..393R,2011A&A...528A..90N}. This sharp period cut-off is considered to be related to the nuclear evolution time-scale of the primary component \citep{2020ApJS..249...18C}.
\subsubsection{Physical parameters of Eclipsing Binaries}\label{ew_param}

\subsubsection*{Through empirical relations}

It was found that absolute parameters of the eclipsing binaries depend on the period and the mass ratios of binary systems \citep{2021ApJS..254...10L}. We calculated the physical parameters of EW stars using \citet{2021ApJS..254...10L} empirical relations given as follows:

\begin{center}
M$_{1}$ = (2.94 $\pm$ 0.21)P + (0.16 $\pm$ 0.08) \\
M$_{2}$ = (0.15 $\pm$ 0.17)P + (0.32 $\pm$ 0.06) \\
R$_{1}$ = (3.62 $\pm$ 0.13)P + (0.04 $\pm$ 0.05) \\
R$_{2}$ = (1.56 $\pm$ 0.13)P + (0.16 $\pm$ 0.05) \\
L$_{1}$ = (13.98 $\pm$ 0.75)P + (3.04 $\pm$ 0.27) \\
L$_{2}$ = (3.66 $\pm$ 0.26)P + (0.69 $\pm$ 0.09) \\
\end{center}
We also calculated the ages of these EW stars using the method described in \citet{2013MNRAS.430.2029Y} and used by \citet{2021ApJS..254...10L}. The secondary components of EW-type binaries are currently less massive than the primary star. However, the secondary stars are found to be more luminous than the luminosity expected for the current mass. This extra luminosity is interpreted as a hint of its mass before mass transfer. The M$_{L}$ is defined as the mass of an isolated star with the luminosity equivalent to the luminosity of the secondary star. The quantity $\delta$M which is equal to M$_{L}$ - M$_{2}$ is used to find the initial mass of the secondary star, M$_{2i}$. The values of M$_{L}$ and M$_{2i}$ are calculated as follows:
\begin{center}
M$_{L}$ = $\Bigg(\frac{L_{2}}{1.49}\Bigg)^{0.237}$ \\
M$_{2i}$ = M$_{2}$ + 2.5($\delta$M - 0.07)$^{0.64}$
\end{center}
The initial mass ratio can be calculated in the form as follows:
\begin{center}
$\frac{1}{q_{i}}$ = $\frac{M_{1} - (M_{2i} - M_{2})(1 - \gamma)}{M_{2i}}$
\end{center}
where $\gamma$ is the ratio of the mass loss from the system to the (M$_{2i}$ - M$_{2}$) value which we considered as 0.664 for these calculations \citep{2021ApJS..254...10L}. The value of mass ratio was found to vary with mass difference $\delta$M for EW type eclipsing binary while there was no correlation for EA type eclipsing binaries \citep{2013MNRAS.430.2029Y}. The correction was applied for mass difference $\delta$M < 0.35 (refer to Figure 7 of \citet{2013MNRAS.430.2029Y}) for EW stars. We calculated $\delta$M  and found that all the EW stars found in our study have $\delta$M > 0.35 so we did not use the corrections. We employed the following relation for the calculation of the ages of binary components:
\begin{center}
t$_{MS}$ = $\frac{10}{M^{4.05}}$(0.0056(M + 3.993)$^{3.16}$ + 0.042)
\end{center}
where M is the mass of the component of which age is to be determined. The age of the binary system itself for EW type star is calculated as age = t$_{MS}$(M$_{2i}$) i.e. the age of the binary system is calculated by putting M = M$_{2i}$ in the above equation. However, if the age of the binary system calculated through this method comes to be greater than the age of the primary star then the age of the primary star should be taken as the age of the binary star system \citep{2021ApJS..254...10L}. 

The primary temperature of EW systems has been found to be correlated with the orbital period \citep{2020MNRAS.493.4045J,2021ApJS..254...10L}. We calculated the primary temperatures of EW stars having periods below 0.5 days using the following relation given by \citet{2021ApJS..254...10L}: 
\begin{center}
T$_{1}$ = -(2752$\pm$320)$\times$P + (6745$\pm$100)
\end{center}
where T$_{1}$ is the temperature of primary and P is the orbital period of the EW binary.
The values of the physical parameters calculated using the empirical relations are given in Table~\ref{ew_par}. We caution about the reliability of the obtained parameters corresponding to the star with ID 600 as parameters of EW stars with periods greater than 0.5 days have been hardly reliable \citep{2021ApJS..254...10L}. The ages of remaining stars are found to be greater than 0.6 Gyr which is consistent with the formation time constraint of $\geq$ 0.6 Gyr for short-period binaries \citep{2020MNRAS.493.2271H}. The obtained temperature of the primary component corresponds to the spectral class G. These temperature values are consistent with the classification of these stars as EW binaries because the spectral classes of both the components of an EW system are generally found to be later than the F class. 
%
\begin{table}\fontsize{6.5}{6.5}\selectfont
\caption{The calculated parameters of eclipsing binaries using empirical relations. The star id is given by ID. The labels M, R, L, and T denotes mass, radius, luminosity, and temperature. The suffix 1 denote primary star and 2 marks secondary star. The age is given in second from last column.}
 \label{ew_par}
\centering
\begin{tabular}{ccccccccc}
\hline
 ID   &  M$_{1}$  &  M$_{2}$  &  R$_{1}$   &  R$_{2}$  &   L$_{1}$   & L$_{2}$& Age&  T1 \\
      &(M$_{\odot}$)&(M$_{\odot}$)& (R$_{\odot}$)      &  ((R$_{\odot}$)    &   (L$_{\odot}$)     &    (L$_{\odot}$)& (Gyr)&    \\
\hline 
     600& 1.963& 0.412& 2.260& 1.117&11.613& 2.934& 0.589& - \\
     732& 1.069& 0.366& 1.160& 0.643& 7.365& 1.822& 0.755& 5894 \\
    1785& 1.561& 0.391& 1.765& 0.904& 9.703& 2.434& 0.647& 5433 \\
    2458& 0.966& 0.361& 1.032& 0.588& 6.871& 1.693& 0.786& 5991 \\
    3005& 1.083& 0.367& 1.177& 0.650& 7.430& 1.839& 0.751& 5881 \\
    3472& 1.303& 0.378& 1.448& 0.767& 8.476& 2.113& 0.697& 5675 \\
    3535& 1.111& 0.369& 1.211& 0.664& 7.561& 1.874& 0.744& 5855 \\
    4829& 1.176& 0.372& 1.291& 0.699& 7.870& 1.955& 0.727& 5794 \\ 
\hline
\end{tabular}
\end{table}

\subsubsection*{Through model fitting}

Eclipsing binaries exhibit variation in brightness with time due to their orbital geometry and alignment from the observer's point of view. We can determine many physical parameters and properties like period, mass, distance, and radii accurately utilizing photometric and spectroscopic methods. The unique features of the evolution and formation of eclipsing binaries can be studied via continuous monitoring. In our work, we identified 10 eclipsing binaries comprising 8 EW-type of eclipsing variables. Out of these eight EW binaries, the V-band photometric light curves of the two EWs are analyzed with the help of the PHOEBE software package. The rest of the identified eclipsing binaries require more photometric observations as some parts of light curves are missing in them. 

\begin{figure*}
\centering
\includegraphics[width=17 cm, height=10 cm]{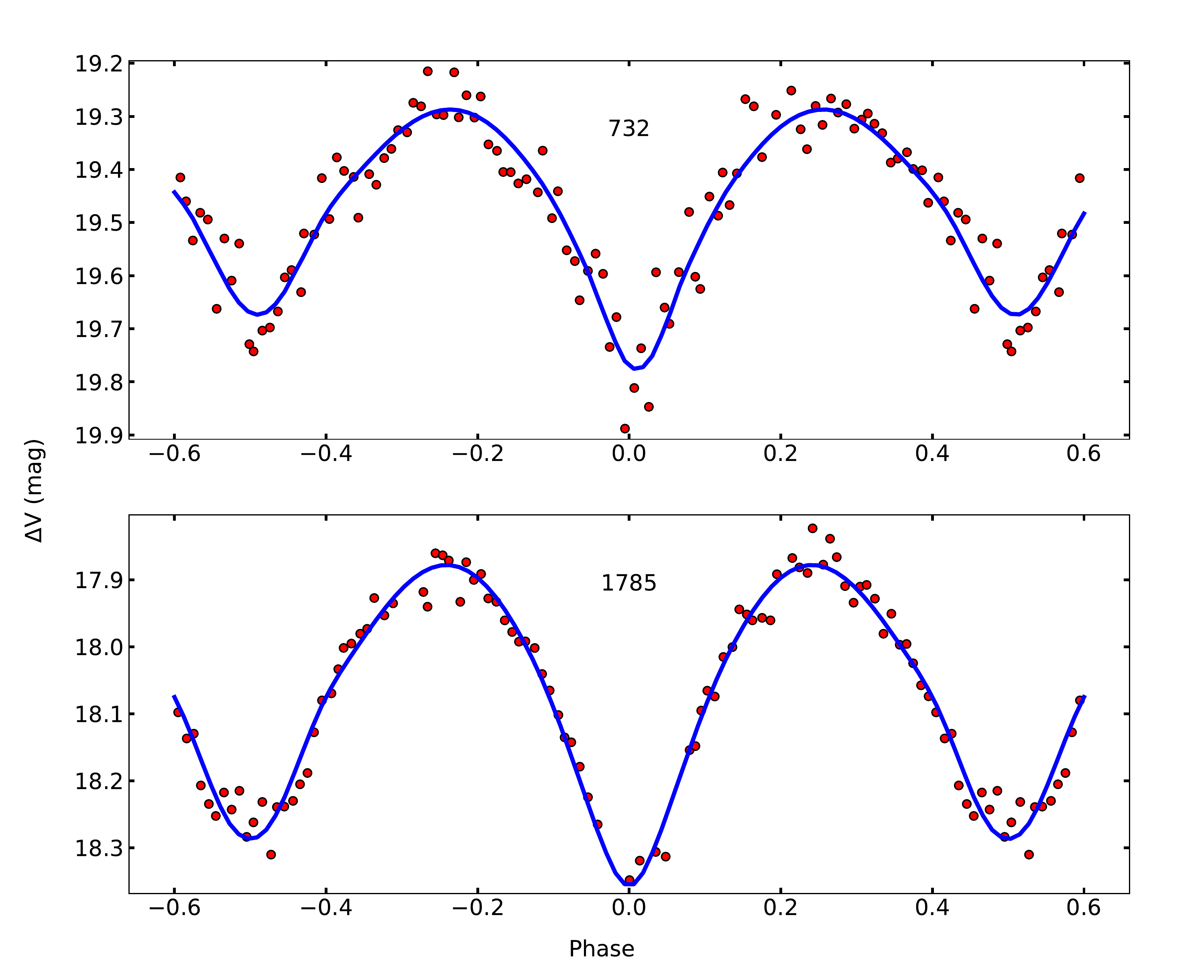}
\caption{The light curves of the two EW binary systems whose physical parameters are estimated using PHOEBE code model fittings. The blue continuous curves show the best model fits of the PHOEBE model fitting code. The IDs for each binary system are given at the top of each subplot. }
\label{ew}
\end{figure*}

PHOEBE is one of the most popular tools used for light curves and radial velocity (RV) curves modeling of eclipsing binaries. It is based on  Wilson–Devinney code which is written in FORTRAN. We used PHOEBE 1.0 which provides a GUI and scripter tool for modeling eclipsing binaries. As the two binaries 0732 and 1785 are reported as EWs, we selected an over-contact binary not in thermal contact as the fitting model in PHOEBE. The temperature of the systems is determined using the color-temperature relation provided by \citep{2010A&ARv..18...67T}. The estimated $T_{eff}$ is used as the temperature of the primary component. The calculation of mass-ratio (q=$m_{2}/m_{1}$) is the first step in modeling the photometric light curve. The accurate determination of the q-parameter requires RV observation but in absence of it, photometric light curves can also be used to constrain the q-parameter around real value. It has been observed that the photometric mass ratio is reliable for total eclipsing binaries only \citep{2013CoSka..43...27H}. So, as the inclination angle decreases the uncertainty in the photometric mass-ratio increases. But due to the absence of RV data, we are dependent only on the q-search method. In this method, some known parameters like the time of minima, period, $T_{eff}$ of primary, surface albedo, and gravity brightening are fixed. The other quantities like inclination (i), $T_{eff}$ of secondary, surface potential ($\Omega$), luminosity ($l_{1}$, and $l_{2}$) are set as free parameters. The q is varied in small steps and corresponding to each q best-fit model is obtained with the number of iterations. The cost function corresponding to the best fit model for each q is noted. The q which corresponds to the minimum cost function is assumed as the actual mass ratio. The process of q-search is repeated for both of the binary systems and the estimated q for 0732 and 1785 are 0.42 and 0.26, respectively. The PHOEBE-scripter was run multiple times with these q-parameters to get the final solution. The fitted light curves are plotted with the observed data in Figure~\ref{ew}.

Both systems 0732 and 1785 show a small difference in the levels of primary and secondary minima, so, some temperature difference between both the components can be expected. The final model shows a temperature difference of almost 500 and 150 for the binaries 0732 and 1785, respectively. The fill-out factor is found to be 0.31 and 0.53 for 0732 and 1785 respectively. The mass and radii of the primary component are determined using the relations by \citep{2009CoAst.159..129G}. The parameters determined from our observations using PHOEBE and these relations are given in Table~\ref{tab:EW_param}.  The system 1785 is also observed by \textit{Zwicky Transient Facility} (ZTF, \citet{2019PASP..131a8002B} survey between 18 May 2018 to 31 May 2022 in zi, zr, and zg band. Most of the photometric observations are in zr-band (1173 data points). To compare the modeling results from our observed data for system 1785, we analyze 1785 ZTF zr-band data with PHOEBE. The \emph{catflags} flag in the ZTF catalog informs user about quality of data \footnote{https://irsa.ipac.caltech.edu/onlinehelp/ztf/help.pdf}. The photometric observations corresponding to \emph{catflags}=0 are used for the modeling. For the ZTF data of the system 1785, the q-search method estimated the q value to be 0.24($\pm$0.03). The secondary temperature and inclination for the system 1785 are determined as 4965($\pm$135) K and 71$^{\circ}(\pm$2), respectively, using the ZTF data. Thus, the modeling results from ZTF data are similar to the results obtained from our observed data.
\begin{table}
\caption{The physical parameters estimated by PHOEBE model fitting on the light curves of the EW binary-systems ID 0732 and 1785.}
\label{tab:EW_param}
\begin{tabular}{c c c c c}
\hline
Parameters     & ID 0732                  &ID 1785   \\
\hline
Period~(days)      &  0.30935                & 0.47661 \\
q                  & 0.42$\pm$0.03           & 0.26$\pm$0.05     \\
i                  & 71$\pm$1                & 72$\pm$5          \\
$T_{1}$ (K)        &  6642                   & 5179           \\
$T_{2}$ (K)        &  6116$\pm$154           & 5003$\pm$55     \\
$\Omega_{1}$       &  2.64$\pm$0.03          & 2.29$\pm$0.02       \\
$\Omega_{2}$       &  $\Omega_{1}$           & $\Omega_{1}$           \\
$L_{1}/(L_{1}+L_{2})$ & 0.752                & 0.794            \\
$M_{1}~(M_\odot)$  & 1.06                    & 1.51              \\
$M_{2}~(M_\odot)$  & 0.44                    & 0.39           \\
$R_{1}~(R_\odot)$  & 1.03                    & 1.65            \\
$R_{2}~(R_\odot)$  & 0.71                    & 0.92           \\
\hline
\end{tabular}
\end{table}

All the results are determined using the photometric mass ratio and V-band photometric data. For the modeling, binned light curves are used and there is also some scattering in data points. These results can be considered preliminary results as more accurate results will require multi-band photometry of these systems. The use of long-term multi-band photometric data can also help in the analysis of other long-term effects like period variation, surface activity, etc. 

%

\subsection{Rotational Variables}
Stars showing variability caused by axial rotation of surface having non-uniform brightness are known as rotational variables. The variation in the surface brightness may be due to the presence of spots and other magnetic activities. The rotational rate of the surface of a star decreases with increasing age and this property is used as an age indicator for individual stars \citep{2007ApJ...669.1167B}. The late-type stars (G or K type) with true color (B-V)$_{0}$ values larger than 0.5 mag \citep{2008ApJ...687.1264M} are classified as rotational variables. In our study, We found total 15 rotational variables having IDs 608, 749, 1155, 1173, 1217, 1399, 1812, 2302, 2618, 2699, 2999, 3717, 3792, 4405, and 5423. The locations of these 15 stars on the CMD were compatible with their classification as rotational variables.

The rotational variable star 2999 is also a member star which gives us opportunity to utilize the period-age relations known to apply for the rotational variables.  We used empirical period-age relations given by \citet{2019AJ....158..173A} to calculate the period for the star 2999.  The G$_{BP}$-G$_{RP}$ value for the star was found to be 1.1254 mag as given in Table~\ref{abs_mag}. We took age of the cluster to be 447 $\pm$ 52 as previously determined by us \citep{2020MNRAS.494.4713M}. Using the period-age relation, the period of the rotational variable 2999 was predicted to be 8.8450 days which is much larger than the period value of 2.93362 days determined in the present study.

Stars 2618 and 5423 locations on the CMD were in the region populated by only rotational ellipsoidal variables. This indicates that these two stars can be classified as rotational ellipsoidal type variables which are close binaries exhibiting variability without eclipses due to distortion of orbital motion by the companion star. These variables are known to have generally smaller amplitude compared to contact binaries \citep{2017MNRAS.469.3688D} which is also the case with stars 2618 and 5423. The stars 1173 and 1812 were located in the region occupied by rotational ellipsoid variables and RS CVn variables on CMD. These two stars did not show distortion of the eclipses in light curves generated from current data to ascertain their classification as RS CVn stars.

\begin{figure}
\includegraphics[width=8 cm, height=8 cm]{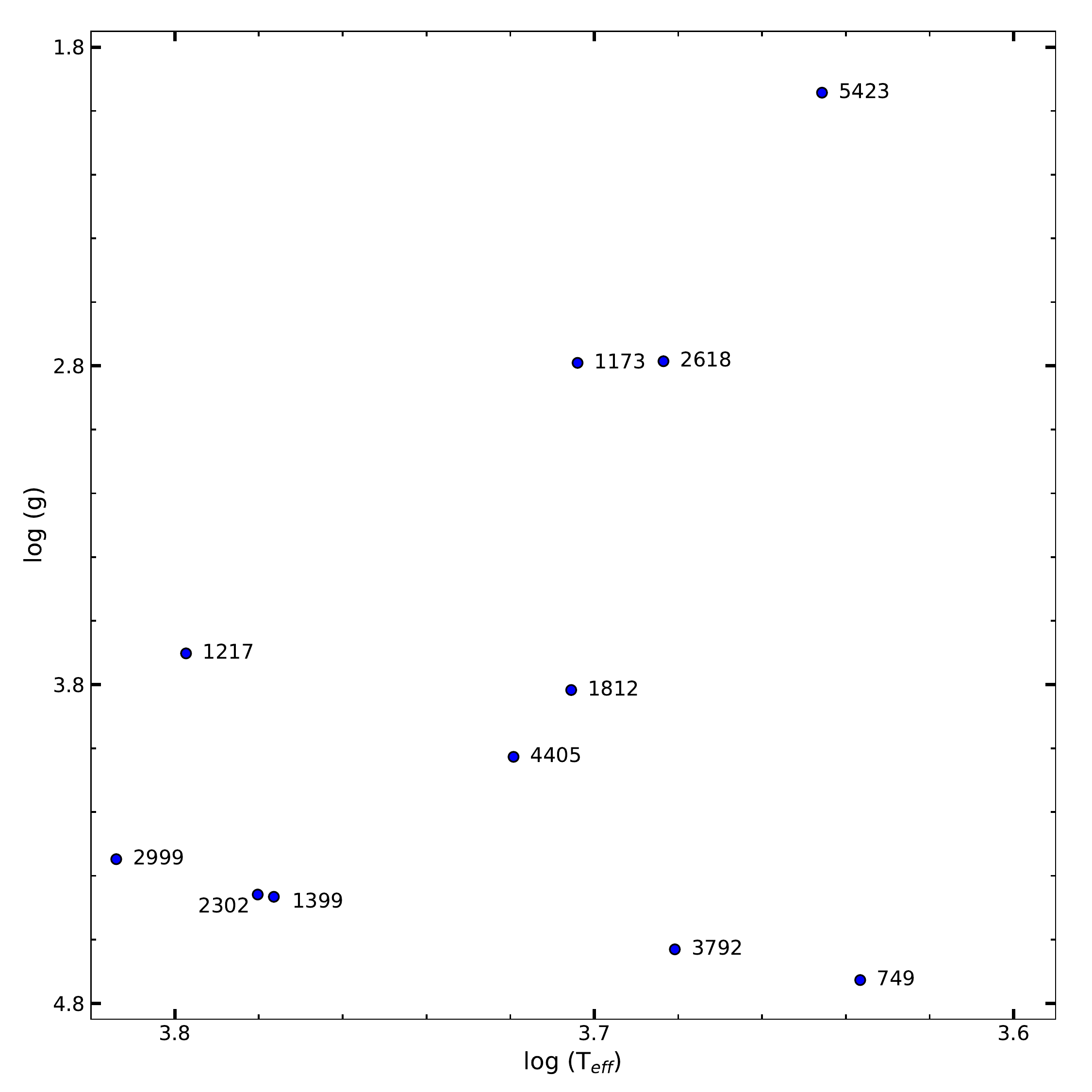}
\caption{Plot of log(g) versus log(T$_{eff}$) diagram for rotational variables whose log(g) and T$_{eff}$ values are available in \citet{2019A&A...628A..94A} and \citet{2019AJ....158...93B} catalogs.}
\label{logt_logg}
\end{figure}

We have shown the distributions of the rotational variables in log (g) versus log (T$_{eff}$) diagram in Figure~\ref{logt_logg}. We noticed that four stars having IDs 1217, 1399, 2302, and 2999 are separated from others in the log (g) versus log (T$_{eff}$) plot. These four stars satisfy the criteria for dwarf stars given by \citet{2011AJ....141..108C}. The magnetic braking causes a decline in the rotation rates of stars with  increasing ages \citep{1972ApJ...171..565S}. The dwarf stars having spectral class later than F have been found to be long-period variables generally with the period of the order of weeks compared to the smaller period of early-type dwarf \citep{2011AJ....141..108C}. The dwarf stars 1217, 1399, 2302, and 2999 also have comparatively larger periods than the periods of other rotating variables as the effective temperatures given in Table~\ref{basic_par} indicate that these are spectral class G stars. The star 2618 has a mass of 1.36 $M_{\odot}$ \citep{2019A&A...628A..94A} and the smallest period of 0.248 days in our list of identified rotational variables. An important transition occurs around 1.3 $M_{\odot}$ stars above which stars are generally fast-rotating. In these stars, convective envelopes become so thin that it becomes unable to produce strong magnetic braking through magnetic winds \citep{2013ApJ...776...67V}.
\subsection{Variability caused by Pulsations}
\subsubsection{$\delta$-Scuti stars}
\begin{figure}
\includegraphics[width=9 cm, height=7 cm]{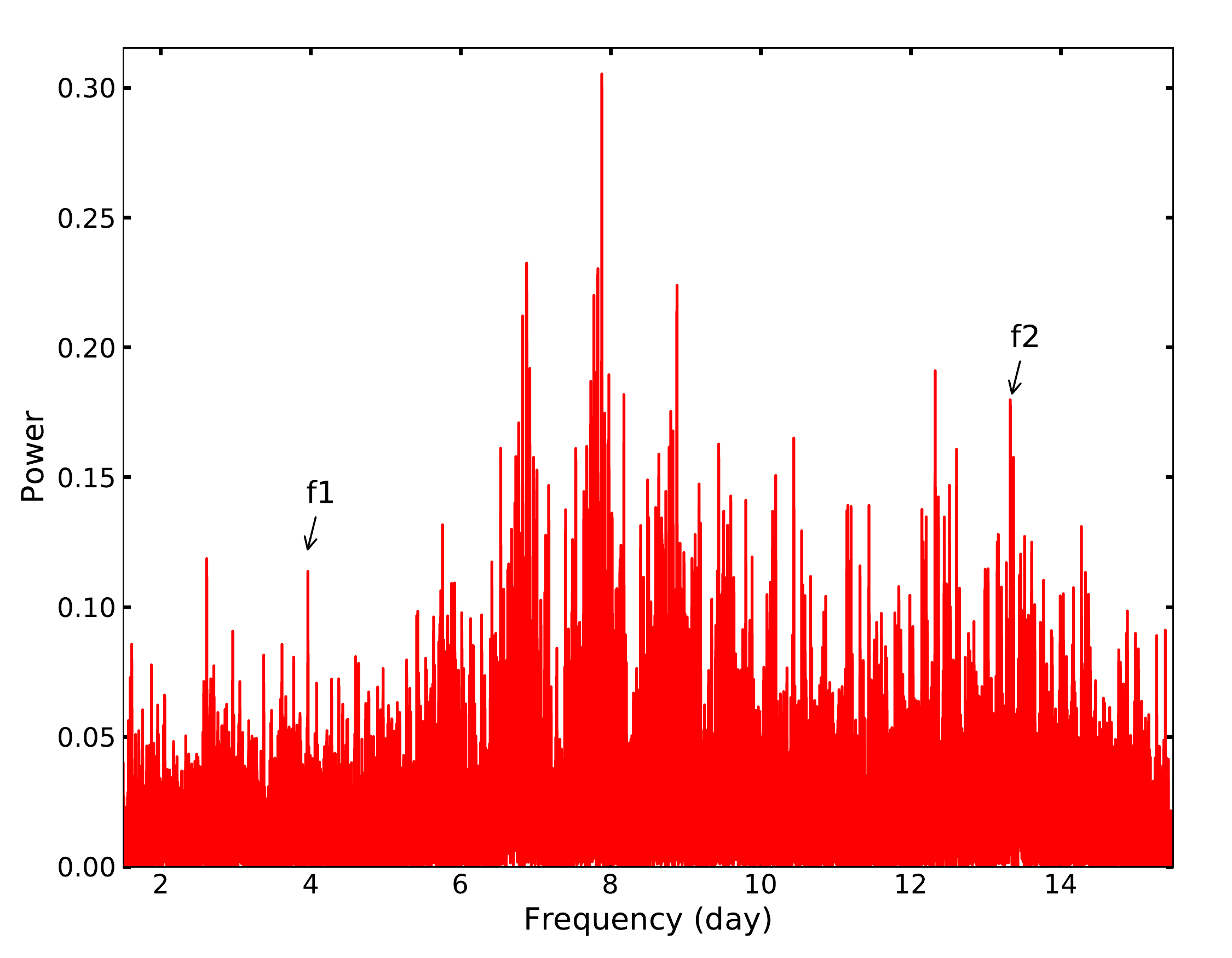}
\caption{Plot of power spectrum for star 2483. The f1 denotes frequency binary system while frequency f2 corresponds to $\delta$-Scuti type variability.}
\label{power_spec_2483}
\end{figure}
$\delta$-Scuti stars are short period pulsating variables found in the theoretical instability strip of $\delta$-Scuti stars on the H-R diagram. These variables have periods less than 0.3 days and amplitude of the order of a few 0.1 mag. $\delta$-Scuti stars are A0-F5 spectral type stars lying on the main sequence or post-main sequence stages of the H-R diagram \citep{2000ASPC..210....3B} driven by $\kappa$ mechanism of pulsation. $\delta$-Scuti stars are intermediate mass (1.5-2.5 M$_{\odot}$) stars \citep{2020Natur.581..147B} with effective temperature in range of 6300-8600 K \citep{2011A&A...534A.125U}. Star 2483 is located in the $\delta$-Scuti instability strip of the H-R diagram for NGC 381. This star exhibits properties of $\delta$-Scuti for a period of 0.07504 days while it also shows the light curve of the binary system for a period of 0.25379 days. The power spectrum for 2483 is shown in Figure~\ref{power_spec_2483}. The amplitude corresponding to the period of 0.07504 days was found to be 0.037 mag as also listed in Table~\ref{basic_par}. The effective temperature and mass of this star are reported to be  6731 K and 1.696 M$_{\odot}$, respectively \citep{2019AJ....158...93B,2019A&A...628A..94A}. On the basis of the period, amplitude, and the location on the H-R diagram star 2483 can be classified as a $\delta$-Scuti variable star. This star is located within the region assigned for $\delta$-Scuti stars on the CMD \citep{2021AJ....162..133S} while it is very close to the constrained region for $\delta$-Scuti stars on the CMD given by \citet{2019A&A...623A.110G}. Therefore, we classified star 2483 as a $\delta$-Scuti variable. This star is an interesting target for further follow-up spectroscopic studies as it seems to be binary star with  a $\delta$-Scuti component. 
\begin{figure}
\includegraphics[width=9 cm, height=5 cm]{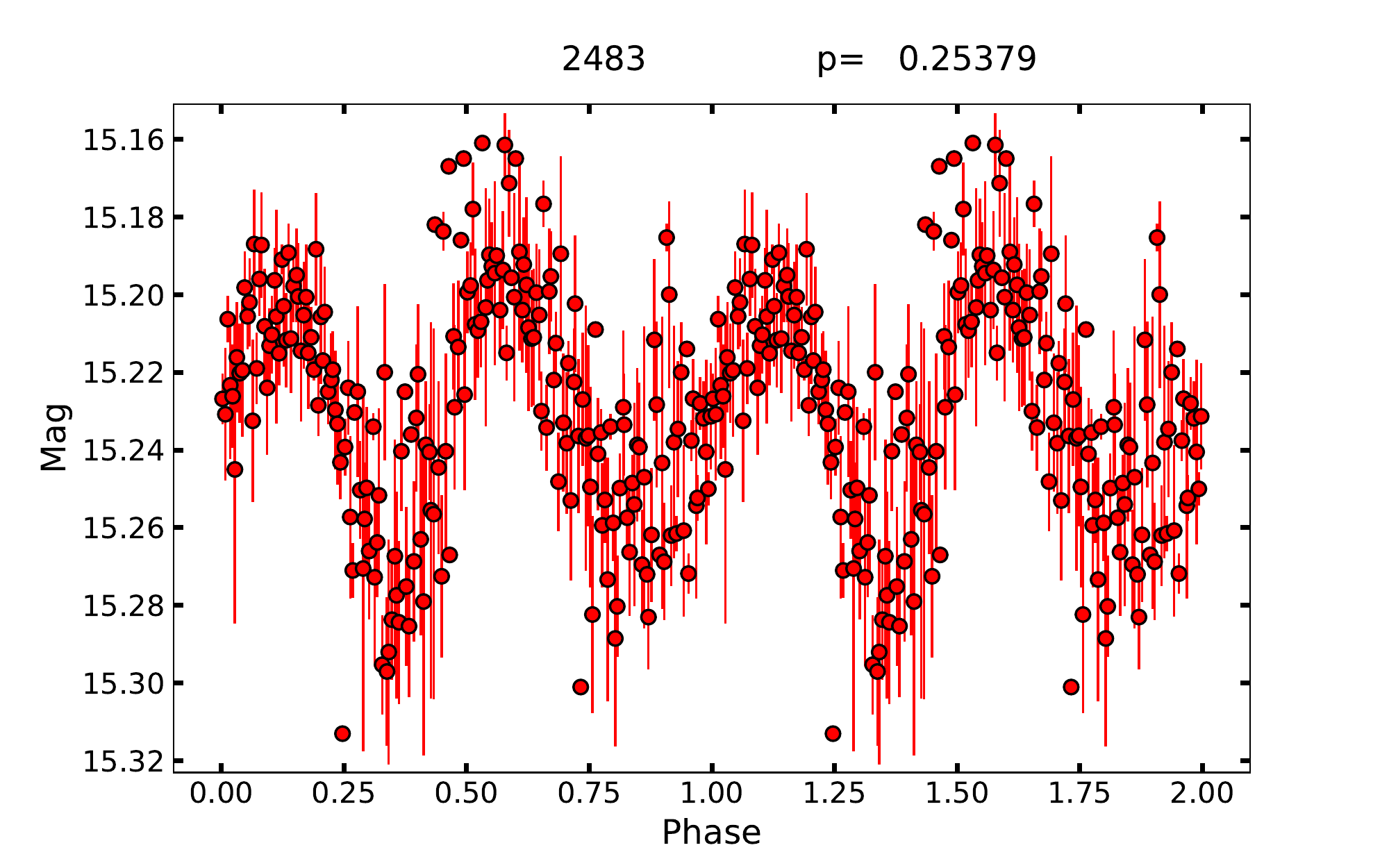}
\caption{Plot illustrating Eclipsing nature of the $\delta$-Scuti star 2483. The light curves is constructed for the period of 0.25379 day.}
\label{eclipse_scuti}
\end{figure}
\subsubsection{$\gamma$ Dor variable stars} 
\begin{figure*}
\includegraphics[width=9 cm, height=6 cm]{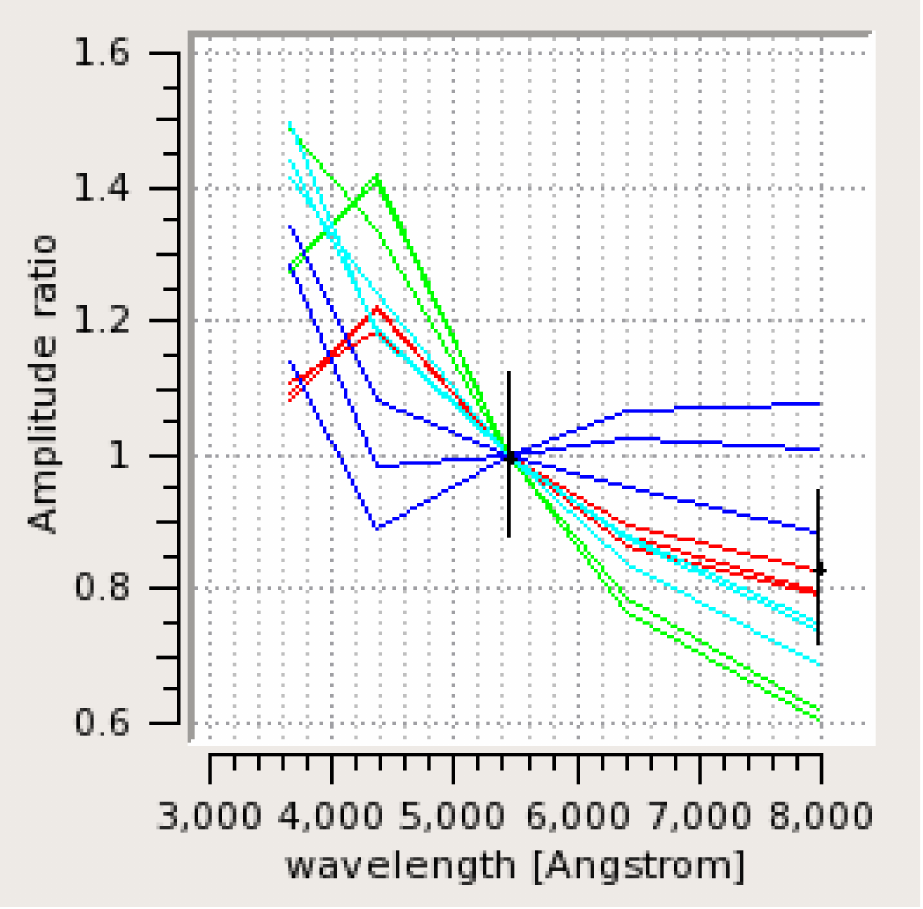}
\includegraphics[width=9 cm, height=6 cm]{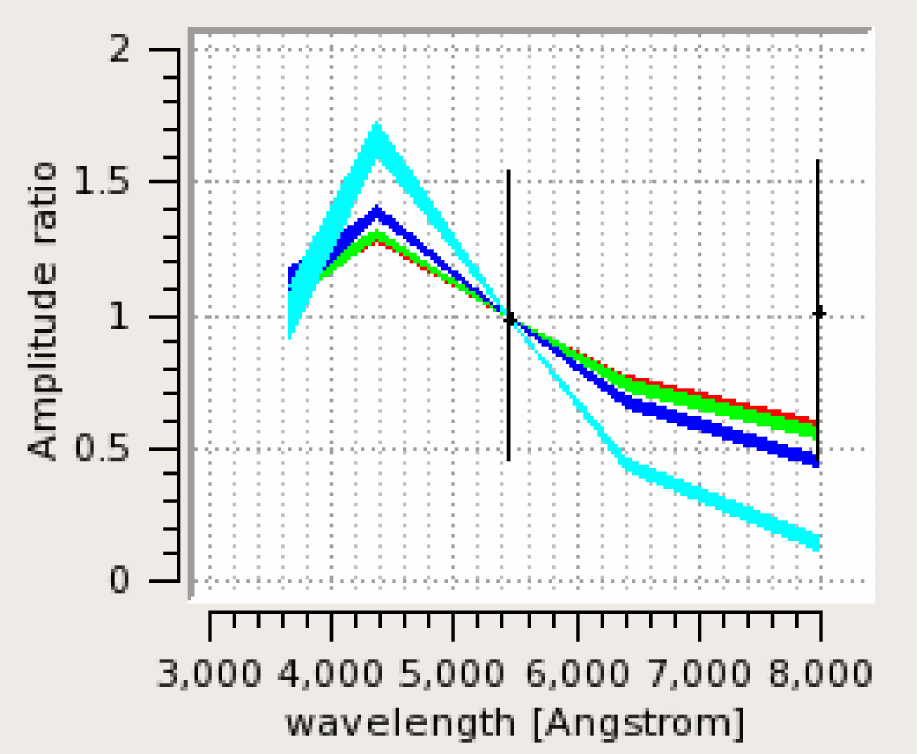}
\caption{The plots of amplitude ratios for mode identification through FAMIAS package using multi-band photometric data. The left and right panel show plots of the amplitude ratios for stars 2483 and 1596, respectively. The mode l=0, 1, 2, and 3 are shown by red, green, blue, and cyan respectively.}
\label{mode}
\end{figure*}
$\gamma$ Dor stars are g-mode pulsating stars with spectral class A7-F5 lying in the $\gamma$ Doradus instability strip in the H-R diagram. The location of the $\gamma$ Dor variables on the H-R diagram is important for understanding the interior of stars as it is the transition zone where convective cores and radiative envelopes energy transfer changes to radiative cores and convective envelopes energy transfer. This property associated with $\gamma$ Dor stars helps in constraining the theoretical models of heat transfer in stars. The pulsation in $\gamma$ Dor stars is non-radial g-modes and pulsation periods of these stars are between 0.3-3.0 days. The typical amplitude of variation for $\gamma$ Dor stars are $\sim$0.1 mag. These stars are less massive than $\delta$ Scuti stars having a mass range from 1.5 M$_{\odot}$ to 1.8 M$_{\odot}$. The variable star with ID 1596 in our list of variables exhibits properties of a $\gamma$ Dor star as suggested by its amplitude, period, and location in the $\gamma$ Doradus instability strip. The location of this star on the CMD was in the region mostly occupied by $\gamma$ Dor stars. The period was found as 0.43500 day for the star 1596. The effective temperature for star 1596 was 6739 K reported by \citet{2019A&A...628A..94A} which is in the range of temperature 6700-7400 K expected for the $\gamma$ Dor stars. Therefore, we classified star 1596 to be $\gamma$ Dor star. The pulsation mode identification of this $\gamma$ Dor star was done using multi-band photometric data through FAMIAS software \citep{2008CoAst.157..387Z} as discussed in the following section.
\subsubsection{Mode identification of pulsation}\label{famias}
We used time-series data obtained in V and I bands for the identification of pulsation modes of $\delta$-Scuti and $\gamma$ Dor stars through FAMIAS software packages \citep{2008CoAst.157..387Z}. The results are compared with prediction of models for l = 0,1,2,3,4. The effective temperatures, surface gravity, and metallicity values required for model fitting in FAMIAS were taken from \citet{2019A&A...628A..94A}. The multi-band photometric mode determination using FAMIAS is based on the relation between amplitude ratios and phase differences obtained for various photometric bands \citep{1988Ap&SS.140..255W}. We identified 2483 as a $\delta$-Scuti star while 1596 was characterized as a $\gamma$ Dor star. The plots of amplitude ratios versus phase differences are given in Figure~\ref{mode} for both the pulsating stars. The $\delta$-Scuti star 2483 was found to be pulsating in l = 0 which corresponds to radial pulsation \citep{2019MNRAS.486.4348Z,2020MNRAS.493.4186J}. We could not find a good fitting for star 1596, however, it seems to be pulsating in l = 0 mode. The low harmonic degree pulsation of l=0 is typical for $\delta$-Scuti and $\gamma$ Dor stars \citep{2017A&A...597A..29S}.
\subsection{Misc Variable}
The variable stars which could not be classified in any one type of variability class were called miscellaneous type variables. A total of 30 variable stars were classified as Miscellaneous in the present study. Stars 2472, 2673, and 5451 lie between SPB and $\delta$-Scuti instability strips. This region in the H-R diagram is prohibited for pulsation according to standard stellar models. \citet{2016A&A...595L...1M} classified these stars as fast-rotating pulsating B (FaRPB) variable stars based on a spectroscopic study which revealed that the majority of these stars were fast-rotating. FaRPB stars are known to be the short period (p<0.55 d) variables \citep{2016A&A...595L...1M,2020MNRAS.499..618J}. However, these variables were classified as miscellaneous variables since they have periods significantly larger than 0.55 days. The light curves of stars 1689 and 1935 show a hint of eclipsing binary but we could not confirm their exact nature and assigned the Misc category to them. Star 2245 was located near the blue edge of the $\delta$-Scuti instability strip. However, its period of 0.96225 days was significantly higher than the period of less than 0.3 days expected for $\delta$-Scuti stars. Star 737 lie close to the blue edge of the $\delta$-Scuti instability strip, however, it has a period of 5.908 days which is very high for any $\delta$-Scuti star.  

\begin{figure*}
\centering
\includegraphics[width=17 cm, height=3.5 cm]{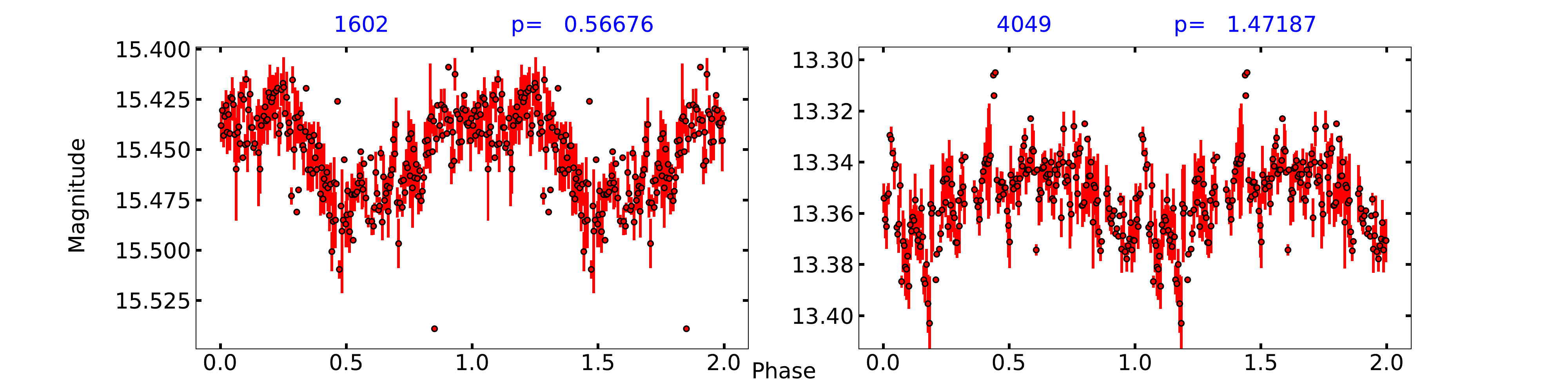}
\includegraphics[width=17 cm, height=3.5 cm]{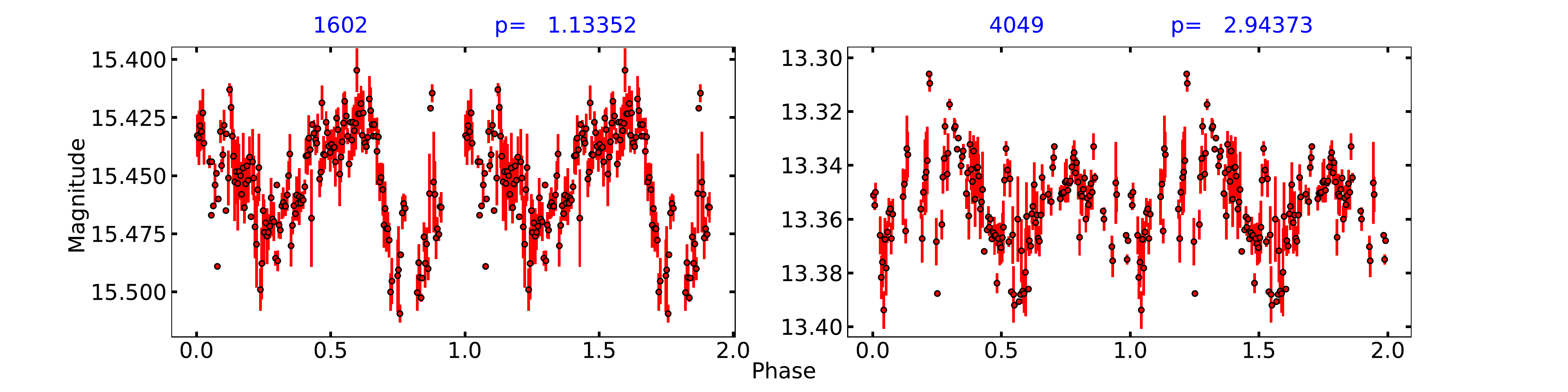}
\caption{Possible RS CVn variable stars. The light curves in upper panel are for the obtained period while light curves in lower panel are plotted for twice the obtained period. The star IDs and the periods are given on the top of each plots.}
\label{rscvn}
\end{figure*}

The stars 1602 and 4049 have been classified as RS CVn variables in the AAVSO International Variable Star Index (VSX) catalog \citep{2006SASS...25...47W}. The RS CVn variables are eclipsing binary systems with the primary star of the spectral class of F-G type. These binary systems are known to have asymmetric light curves caused by chromospheric activities and rotation. The distortion in the light curves is thought to be caused by the uneven distribution of the surface cool spots \citep{2007PASP..119..259E,1979ApJ...227..907E}. RS CVn variables show spectral emission lines of Ca II H and K \citep{2008A&A...487..709Z}. The light curves of these stars are the results of spot modulation of the eclipses. The LCs for these two variable stars are shown in Figure~\ref{rscvn}. As it is clear from the figure that the light curves of these stars do not exhibit clear eclipses so these stars could be rotational or non-eclipsing binaries with large spots on the one or both of the components. However, the effective temperatures of the stars 1602 and 4049 were reported to be 6922 and 7988 K by \citet{2019AJ....158...93B} indicating that these systems are earlier than K spectral type  which is a potential spectral class for RS CVn stars \citep{2020ApJS..249...18C}. The locations of stars 1602 and 4049 on the CMD are compatible with their classification as RS CVn variables. The RS CVn stars with larger (G$_{BP}$ - G$_{RP}$) color values generally have greater absolute G magnitudes \citep{2019A&A...623A.110G}. This tendency is also found to be true in the case of RS CVn stars identified in the present study as can be seen in Table~\ref{abs_mag}. Considering all these properties of the stars 1602 and 4049 we classify them as Miscellaneous variables.

\subsection{Comparison}
We searched variable stars identified by us in the previously available variable stars catalogs. We found 11 out of 57 variable stars identified by us to be reported in VSX catalog. There were also eight variable stars common with the catalog of periodic variable stars provided by \citet{2020ApJS..249...18C}  using the ZTF data. The periods and classifications of these stars are given in Table~\ref{comp} for comparison. The periods determined in zg and zr bands of ZTF were found to be differ significantly in \citet{2020ApJS..249...18C} catalog for few stars. The observed data points in zr band were more than twice of the data points in zg band for these eight stars so we mentioned only periods in zr band  in Table~\ref{comp}. We found that the classifications of the variable stars given in these catalogs are generally in agreement with the our classifications. We could not find evidence for the variables 1602 and 4049 to be RS CVn stars as they are reported in these catalogs. The periods given in the catalogs are mostly in agreement with the periods determined by us. The remaining 46 variables are not listed in any variable stars catalogs so they can be considered as new variable stars identified by us.
\begin{table}\fontsize{7}{7}\selectfont
\label{comp}
\caption{The comparison of periods and variability type of variable stars reported in the VSX catalog and the ZTF catalog of periodic variable stars provided by \citet{2020ApJS..249...18C} .  The periods and variability types from the VSX and the ZTF catalogs are denoted by prefixes VSX and ZTF. The periods and variability types determined by us are given in the second last and the last columns, respectively.}
\label{comp}
\centering
\begin{tabular}{cccccccccccc}
\hline
 ID   &  P$_{VSX}$ & Type$_{VSX}$&  P$_{ZTF}$ & Type$_{ZTF}$ &   P    &   Type  \\
      &  (day)     &              &   (day)&     \\
\hline
    
     298& -              &            EA& 8.54994& EA& 2.88585& EA \\
     732&   0.30936&           EW&             -&     -&0.30934& EW \\
    1397&   2.48934&  EA/DM&            -&    -&2.48932& EA \\
    1602&   0.53806&  RS CVn& 0.53806&  RS CVn&1.13352& Misc \\
    1785&   0.47660&         EW& 0.47660& EW&0.47661& EW \\
    2458&   0.27406&         EW& 0.27405& EW&0.27405& EW \\
    2483&   0.07504&   $\delta $-Scuti&            -&      -&0.07504& $\delta$-Scuti\\
    3005&   0.31404&        EW& 0.31404& EW&0.31405& EW\\
    3472&   0.38883&        EW& 0.38883& EW&0.38882& EW \\
    3535&   0.31240&        EW& 0.32340& EW&0.32339& EW \\
    4049&   9.26598& RS CVn& 9.26598& RS CVn& 2.94373& Misc\\
\hline
\end{tabular}
\end{table}

\begin{figure*}
\includegraphics[width=8 cm, height=5 cm]{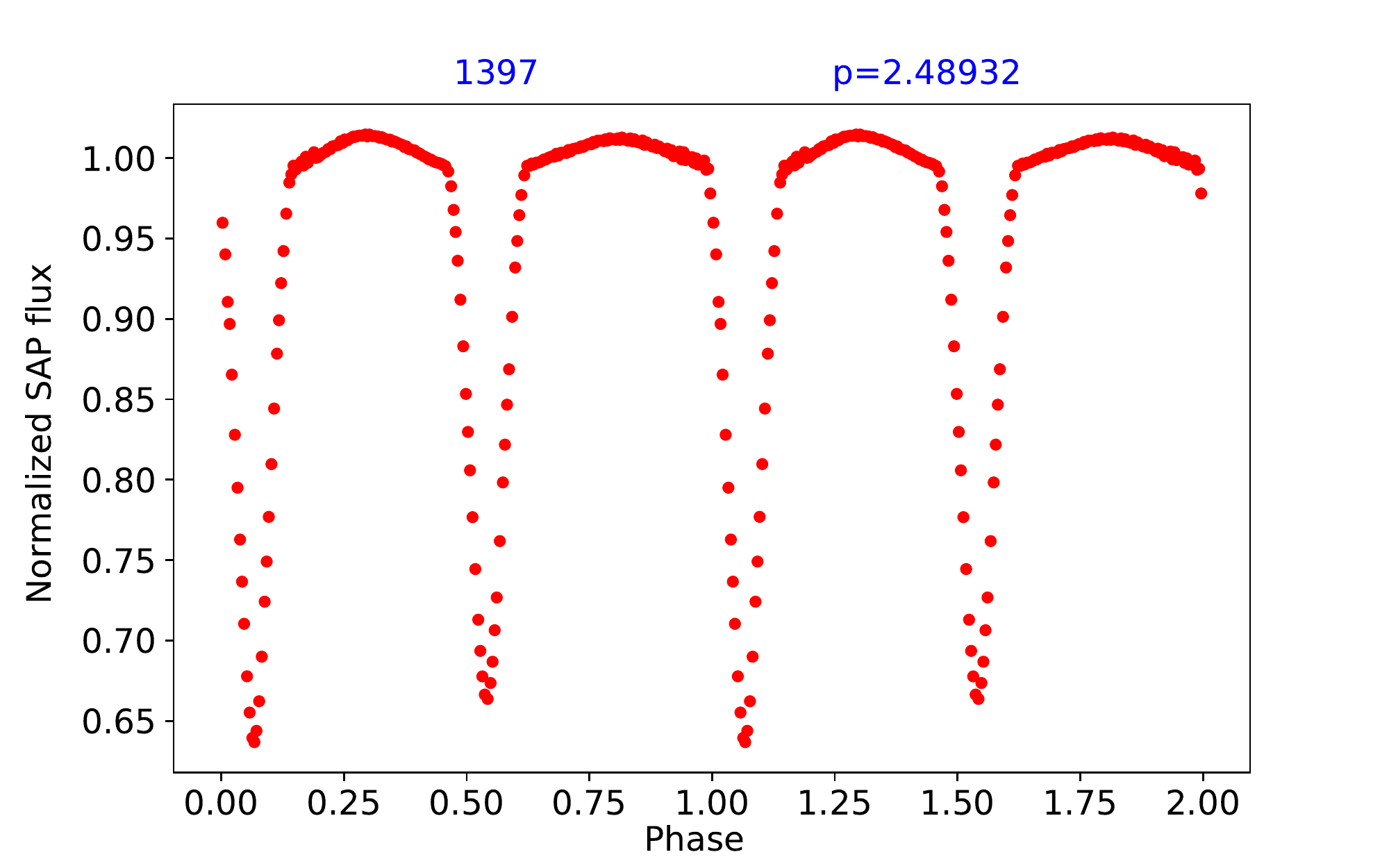}
\includegraphics[width=8 cm, height=5 cm]{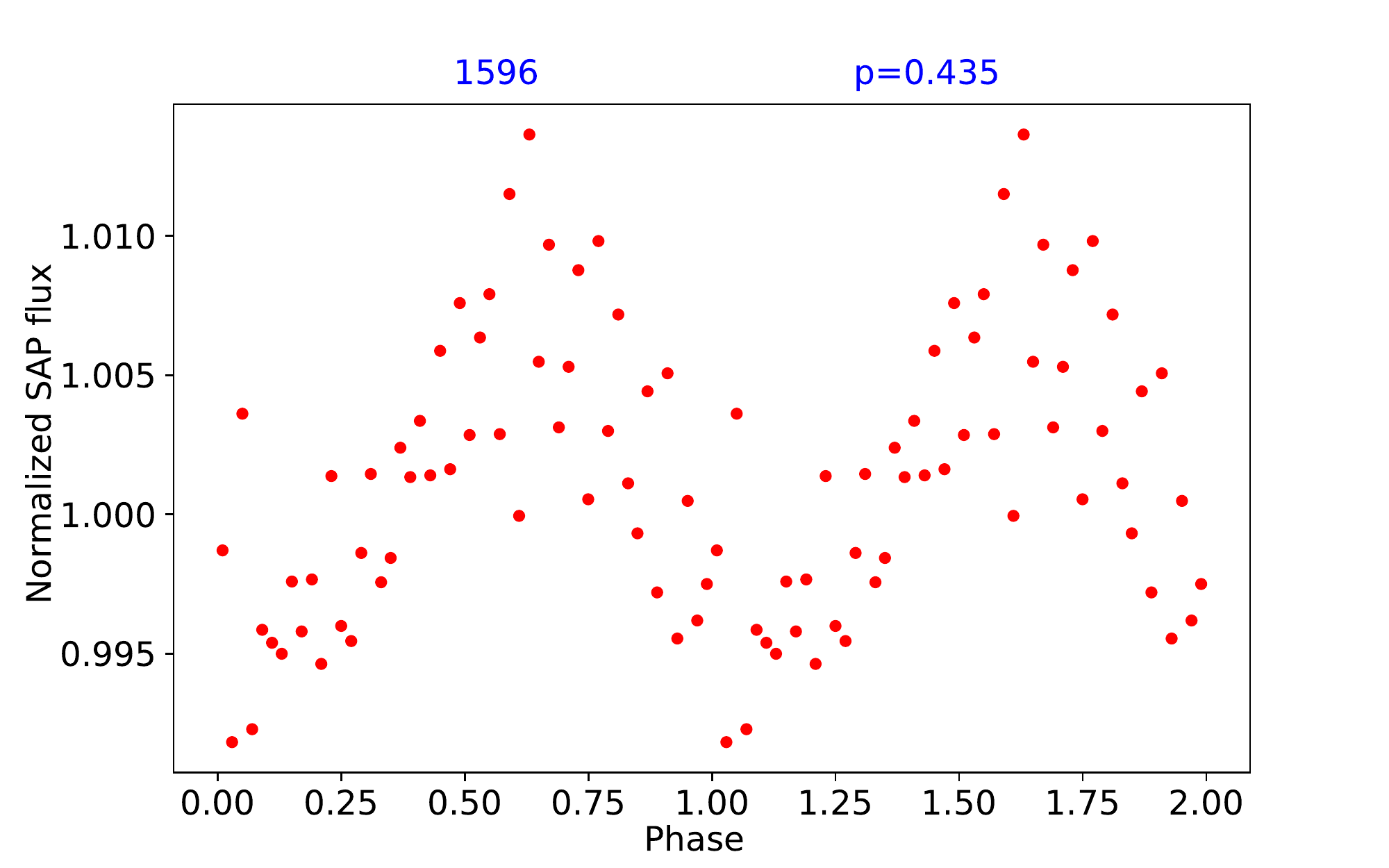}
\includegraphics[width=8 cm, height=5 cm]{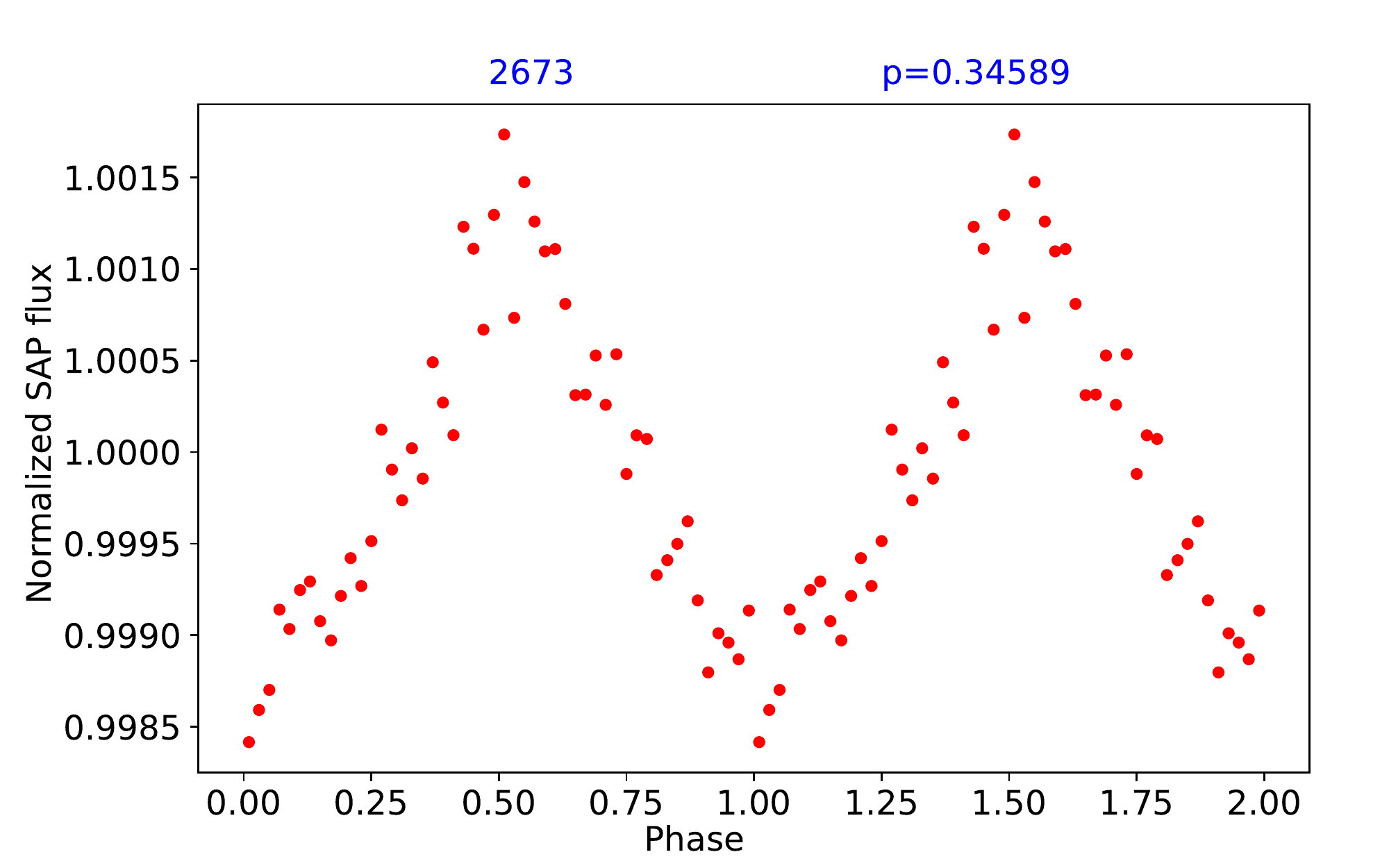}
\hspace{1.5 cm}
\includegraphics[width=8 cm, height=5 cm]{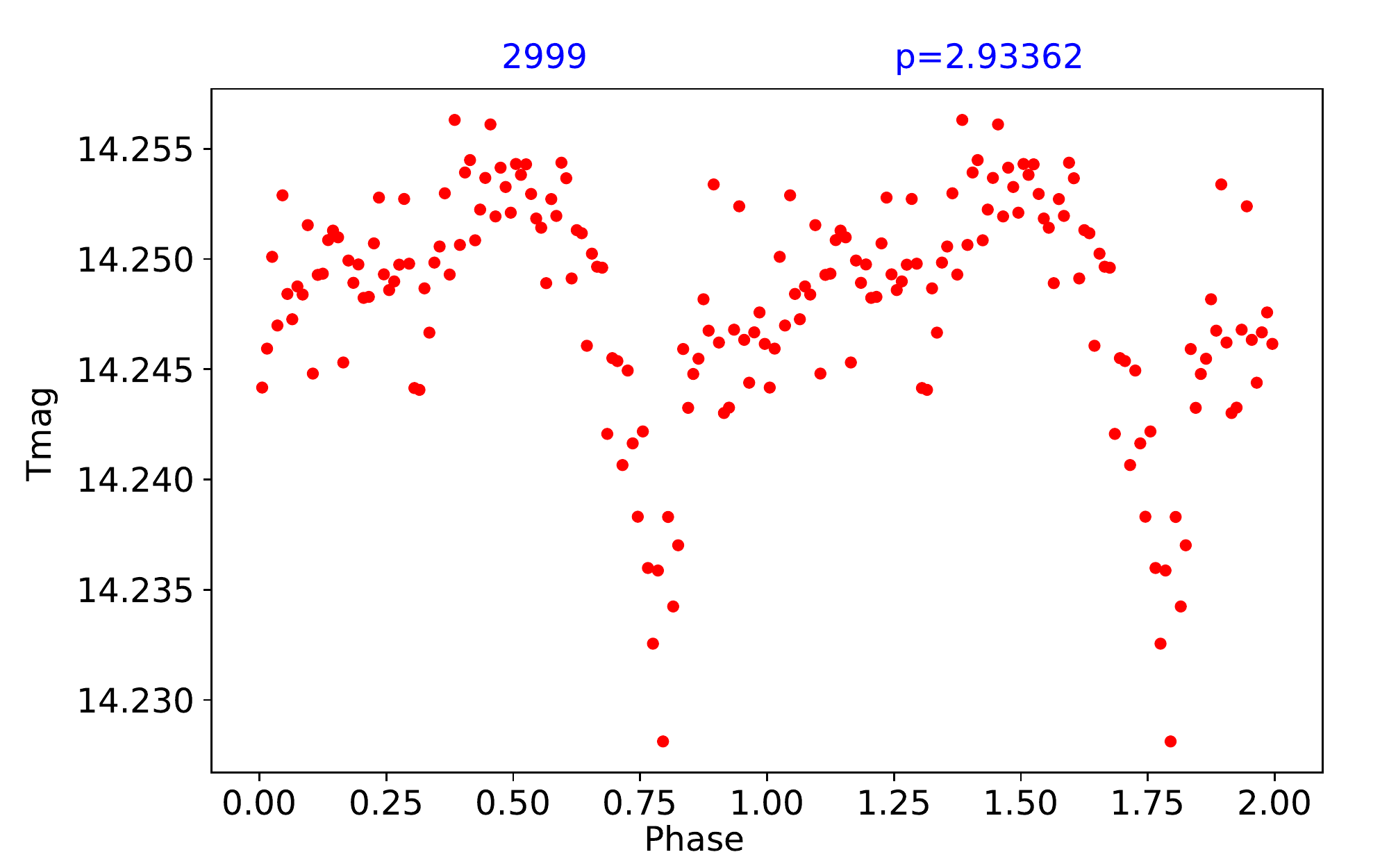}
\caption{The light curves for variable stars found in TESS archives. The star IDs and periods are denoted at the top of each plot. The TESS mag (Tmag) for the star ID 2999 is calculated from the PSF flux provided in the TESS archives using the relations given by \citep{2021AJ....162..170H}.}
\label{lc_tess}
\end{figure*}

We also check Transiting Exoplanet Survey Satellite (TESS) archive to compare our observations. We found six of our detected variable stars namely 1397, 1596, 2302, 2472, 2673, and 4049 in TESS archives having Simple Aperture Photometry (SAP) fluxes estimated using QLP pipelines. The cadence for these stars is 1800 seconds in the archived data. We could extract very nice light curve for the star 1397 which is a bright star with TESS magnitude (Tmag) $\sim$ 9.7 mag \citep{2022yCat.4039....0P} showing periodicity of amplitude $\sim$0.3 mag (see Figure~\ref{lc_tess}). The TESS data for stars 1596 and 2673 also produced good light curves, however, the period for 2673 was different than ours. The time-series PSF fluxes of the stars 2999, 4361, and 5091  extracted using a PSF-based Approach to TESS High quality data Of Stellar cluster (PATHOS) pipeline were also available in TESS archive. The light curve generated using TESS data for star 2999 was consistent with period obtained from the our observed data. We could not extract any signal of variability for remaining five stars probably due to presence of systematic artifacts in SAP fluxes compared to Pre-search Data Conditioning SAP (PDCSAP) fluxes \citep{2019ApJ...873...97L}.
\section{Summary and Conclusion}\label{discussion}
We extensively observed the open cluster region NGC 381 in V and I bands for the variability study. The region was observed using 1.3-m DFOT telescope equipped with 2k$\times$2k CCD having a large field of view of $\sim18^{'}$ $\times$ $18^{'}$ suitable for variability search. The data were collected on 27 nights over a span of more than one year from 01 October 2017 to 14 January 2019. The data was reduced using PSF technique and standard magnitudes were obtained. In the present study, we identified a total of 57 periodic variable stars in the cluster region NGC 381. These variable stars were classified on the basis of the shape of the light curves, period, amplitude, and location on HR diagrams. We found one $\delta$-Scuti star and one $\gamma$ Dor star. The pulsation modes of these variable stars were examined through FAMIAS package using Johnson V and Cousin I band data and the $\delta$-Scuti star was found to be pulsating in possibly the first overtone mode. To detect eclipsing binaries, we checked the light curves of all the periodic variables for twice the period obtained. The stars were qualified as eclipsing binaries only after distinct identification of primary and secondary minima. This way we found a total of 10 eclipsing binaries comprising 2 eclipsing binaries of Algol type and 8 W UMa type eclipsing binaries. The physical parameters of the EW type binaries were estimated using empirical relations and PHOEBE model fitting packages. We classified late-type variables as rotational variables if the color (B-V)$_{0}$ was found to be larger than 0.5 mag. We identified total 15 rotational variables in the present study. The rotational variability is thought to be present due to the uneven distribution of magnetic cool spots on the surface of the stars. There were also some variables that could not be classified in any particular type of the variables and we classified them as miscellaneous type variables. A total of 30 variables were classified in this class. However, more time-series data complemented with the spectroscopic study is necessary to characterize these variables. 

\section*{Acknowledgements}
This work presents results from the European Space Agency (ESA) space mission Gaia. Gaia data are being processed by the Gaia Data Processing and Analysis Consortium (DPAC). Funding for the DPAC is provided by national institutions, in particular the institutions participating in the Gaia MultiLateral Agreement (MLA). The Gaia mission website is https://www.cosmos.esa.int/gaia. The Gaia archive website is https://archives.esac.esa.int/gaia. This paper includes data collected by the TESS mission, which are publicly available from the Mikulski Archive for Space Telescopes (MAST) at the Space Telescope Science Institute (STScI). Funding for the TESS mission is provided by the NASA's Science Mission Directorate. Based on observations obtained with
the Samuel Oschin 48-inch and the 60-inch Telescope at the Palomar Observatory as part of the Zwicky Transient Facility project. ZTF is supported by the National Science Foundation under Grant No. AST1440341 and AST-2034437 and a collaboration including Caltech, IPAC, the Weizmann Institute for Science, the Oskar Klein Center at Stockholm University, the University of Maryland, the University of Washington, Deutsches Elektronen-Synchrotron and Humboldt University, Los Alamos National Laboratories, the TANGO Consortium of Taiwan, the University of Wisconsin at Milwaukee, Trinity College Dublin, Lawrence Livermore National Laboratories, Lawrence Berkeley National Laboratories, and IN2P3, France. Operations are
conducted by COO, IPAC, and UW.
\software{IRAF \citep{1986SPIE..627..733T,1993ASPC...52..173T}, DAOPHOT II \citep{1990ASPC....8..289S}, PHOEBE \citep{2005ApJ...628..426P}, FAMIAS \citep{2008CoAst.157..387Z}.}\\

\appendix 

\begin{figure*}
\centering
\includegraphics[width=19 cm, height=14 cm]{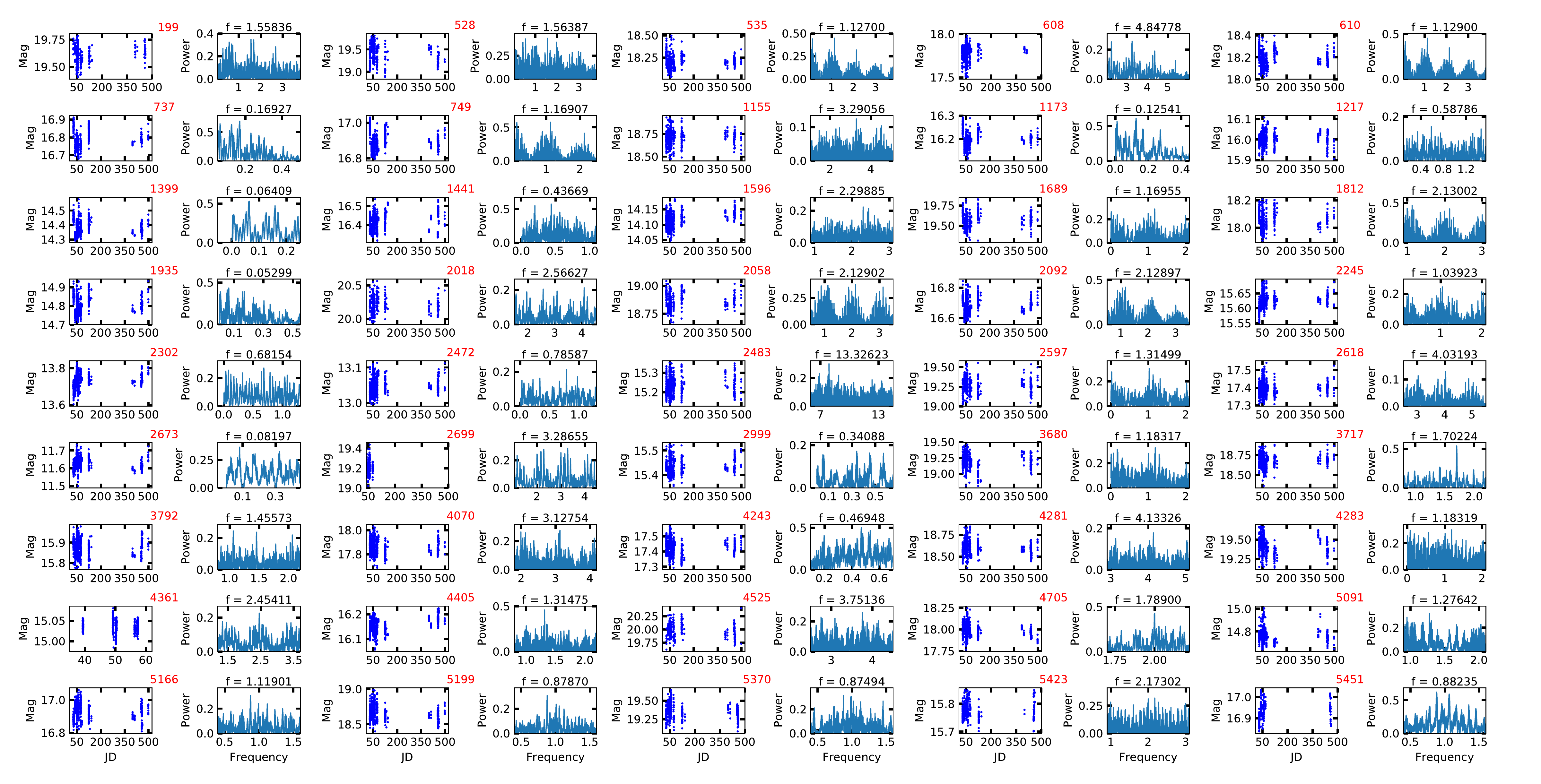}
\caption{The mag versus JD and Lomb-Scargle periodogram plots for the 45 variable stars whose phase folded light curves are shown in Figure~\ref{lc_var}. The star IDs and phase folding frequencies are given at the top of the subplots.}
\label{periodogram}
\end{figure*}
%
\bibliographystyle{aasjournal}
\bibliography{main}{}

\end{document}